\documentclass[11pt]{article}
\usepackage{cite}
\usepackage{amsfonts,amsmath,amssymb}
\usepackage{enumerate}
\usepackage{hyperref}
\usepackage{bbm}
\usepackage{nicefrac}
\usepackage[all]{xy}
\usepackage{graphicx}
\usepackage{bm}
\usepackage{makecell}
\usepackage[table]{xcolor}
\usepackage{longtable}		
\usepackage{subcaption}		
\usepackage{tikz}			
\usepackage{arydshln}

\usepackage{upgreek}

\usepackage{tocloft}			

\usepackage{float}			
\usepackage{lscape}                 

\usepackage{booktabs}

\newcommand{\listnotetitle}{\Large List of Notes}
\newlistof{notes}{nt}{\listnotetitle}

\def\sst#1{{\scriptscriptstyle #1}}

\def\0{{\sst{(0)}}}
\def\1{{\sst{(1)}}}
\def\2{{\sst{(2)}}}
\def\3{{\sst{(3)}}}
\def\4{{\sst{(4)}}}
\def\5{{\sst{(5)}}}
\def\6{{\sst{(6)}}}
\def\7{{\sst{(7)}}}
\def\8{{\sst{(8)}}}

\newcommand{\SO}{\textrm{SO}}
\newcommand{\SL}{\textrm{SL}}

\renewcommand{\i}{\textrm{i}}
\renewcommand{\j}{\textrm{j}}

\renewcommand{\l}{\textrm{l}}
\newcommand{\m}{\textrm{m}}

\newcommand{\vol}{\textrm{vol}}

\allowdisplaybreaks  

\usepackage{multirow}
\usepackage{rotating}

\allowdisplaybreaks

\newlength\Colsep
\begin{document}

\makeatletter
\renewcommand{\theequation}{\thesection.\arabic{equation}}
\@addtoreset{equation}{section}
\makeatother

\begin{titlepage}

\begin{flushright}

MI-HET-822 \\
\today
\end{flushright}

\vspace{25pt}

   \begin{center}
   \baselineskip=16pt

   \begin{Large}

\mbox{ \bfseries \boldmath  Singular limits in STU supergravity}
   \end{Large}

\vspace{25pt}

{\large  Gabriel Larios$^{1}$\,, Christopher N. Pope$^{1,2}$ \,and\, Haoyu Zhang$^{1}$}
		
\vspace{25pt}

	\begin{small}

	{\it $^{1}$ Mitchell Institute for Fundamental Physics and Astronomy, \\
	Texas A\&M University, College Station, TX, 77843, USA}   \\

	\vspace{10pt}
	
	{\it $^{2}$ DAMTP, Centre for Mathematical Sciences, Cambridge University, \\
	Wilberforce Road, Cambridge CB3 OWA, UK}     \\
		
	\end{small}

\vskip 50pt

\end{center}

\begin{center}
\textbf{Abstract}
\end{center}

\begin{quote}

We analyse the STU sectors of the four-dimensional maximal gauged supergravities with gauge groups ${\rm SO(8)}$, $\SO(6)\ltimes\mathbb{R}^{12}$ and $[\SO(6)\times\SO(2)]\ltimes\mathbb{R}^{12}$, and construct new domain-wall black-hole solutions in $D=4$. The consistent Kaluza-Klein embedding of these theories is obtained using the techniques of Exceptional Field Theory combined with the 4$d$ tensor hierarchies, and their respective uplifts into $D=11$ and type IIB supergravities are connected through singular limits that relate the different gaugings.

\end{quote}

\vfill

\thispagestyle{empty}

\end{titlepage}

\tableofcontents


\section{Introduction}

	On certain backgrounds, string theory admits a consistent truncation to a subset of modes in its Kaluza-Klein (KK) spectrum whose complete non-linear dynamics is captured by a gauged supergravity in lower dimensions. Notable examples include the reduction of eleven-dimensional supergravity on ${\rm AdS}_4\times S^7$ \cite{Duff:1983gq,deWit:1984nz,deWit:1986oxb,Nicolai:2011cy} and on ${\rm AdS}_7\times S^4$ \cite{Pilch:1984xy,Nastase:1999cb,Nastase:1999kf}, and the reduction of type IIB supergravity on ${\rm AdS}_5\times S^5$ \cite{Kim:1985ez,Cvetic:2000nc,Cvetic:2000dm}. 
	
	Such consistent truncations are in general difficult to construct, but when they exist they constitute a fundamental tool for obtaining solutions in ten and eleven dimensions, as the consistency of the truncation guarantees that every solution of the lower-dimensional gauged supergravity can be embedded into a configuration that solves the equations of motion of the parent theory. This approach has been particularly fruitful in holographic set-ups, where the gauged supergravity techniques have made possible the construction of hundreds of different AdS solutions -- see \cite{Comsa:2019rcz,Bobev:2020ttg,Bobev:2020qev,Berman:2021ynm,Berman:2022jqn} for recent surveys -- that are dual to different conformal field theories (CFTs) and can be used, among other things, as a playground to check Swampland conjectures, such as the AdS conjecture \cite{Ooguri:2016pdq} or the CFT distance conjecture \cite{Perlmutter:2020buo}. Furthermore, the relative simplicity of the lower-dimensional theories has allowed the analysis of other holographically relevant phenomena, such as black holes \cite{Duff:1999gh,Cvetic:1999xp} describing finite temperature states in the CFT, or domain walls realising CFT interfaces \cite{Clark:2005te,DHoker:2009lky} and RG-flows between different CFTs \cite{Ahn:2000aq,Bobev:2009ms}. Additionally, the existence of these consistent truncations also facilitates even some intrinsically higher-dimensional computations such as the spectrum of masses of the KK modes \cite{Malek:2019eaz}.
		
	Recently, reformulations of string theory based on its duality symmetries \cite{Coimbra:2012af,Hohm:2013pua} have played a pivotal r\^ole in our understanding of consistent truncations \cite{Lee:2014mla,Hohm:2014qga}. In fact, at present there are theorems \cite{Cassani:2019vcl} that guarantee the existence of these reductions for wide classes of theories and backgrounds in terms of suitable factorisations of duality-covariant fields. However, explicit KK Ans\"atze for the standard higher-dimensional metric and fluxes keeping the entire dependence on the lower-dimensional fields and their derivatives are often not known due to the intricate nature of the dictionary relating the original supergravities to the duality-covariant counterparts. 	
A convenient way to describe the embedding of these non-trivial profiles makes use of the tensor hierarchy, which is a supplement to the $p$-forms in the ungauged supergravity that are introduced so as to guarantee that the gauged theory is formally covariant under the original global symmetry group \cite{deWit:2008gc,deWit:2008ta,Bergshoeff:2009ph,Guarino:2015qaa}. The redundancies so introduced can be eliminated at the level of the field strengths, and an explicit KK Ansatz in terms of the original fields and their derivatives can thus be obtained \cite{Varela:2015ywx,Guarino:2015vca,Larios:2019kbw}. These extra forms allow us to trade some complicated dualisations with respect to the internal metric for much simpler dualisations with respect to the four-dimensional metric.
	
	It has been observed \cite{Catino:2013ppa,Berman:2022jqn} that there exists a non-trivial network of gauged supergravitites in four dimensions that are connected by singular limits of their moduli. When the original gauged supergravity admits an uplift to string theory, one can follow this limit also in higher dimensions \cite{Inverso:2017lrz,DallAgata:2019klf}, and thereby connect one consistent truncation to another. This technique can also be employed to construct new solutions from known ones \cite{Cvetic:1999pu}. In this paper, we employ these relations to describe the truncation of type IIB supergravity on ${\rm AdS}_4\times S^5\times S^1$. There are in fact several inequivalent truncations on this geometry, and in this work we focus on the ones in which the resulting gaugings are $\SO(6)\ltimes\mathbb{R}^{12}$ \cite{Cordaro:1998tx} and $[\SO(6)\times\SO(2)]\ltimes\mathbb{R}^{12}$ \cite{DallAgata:2011aa,DallAgata:2014tph}. The latter can be understood as a modification of the former where an extra vector is dyonically coupled to the matter fields. 
	Even though the consistency of these truncations is well known \cite{Inverso:2016eet}, complete KK Ans\"atze are unavailable. 
	For simplicity, we focus here on the truncation of these four-dimensional theories down to the STU sector so that only six scalars and four vectors can have a non-trivial profile. The STU model obtained from the $\SO(6)\ltimes\mathbb{R}^{12}$ theory can be obtained as a singular limit of the STU model corresponding to the SO(8) gauging of \cite{Cvetic:1999xp}, and we show that the $S^5\times S^1$ background in type IIB can be obtained as a singular limit of eleven-dimensional supergravity on the $S^7$ followed by a circle reduction and T-duality. The explicit Ansatz embedding both theories in type IIB is presented in equations \eqref{eq: defS5xS1z}--\eqref{eq: fiveformSO6SO2}.

	The rest of the paper is structured as follows. In the next section we introduce the 4$d$ models which we are going to analyse, and discuss some of their properties and their reformulation in terms of the tensor hierarchy. Section~\ref{sec: E77ExFT} briefly reviews E$_{7(7)}$-Exceptional Field Theory (ExFT) \cite{Hohm:2013uia} and the description of consistent truncations down to maximal gauged supergravities in the language of generalised Scherk-Schwarz reductions \cite{Hohm:2014qga}, which will be subsequently used to describe the uplift of the STU models into M-theory on $S^7$ and type IIB supergravity on $S^5\times S^1$, and the consistency of these embedding is explicitly checked in a simpler sub-truncation. New solutions in these gaugings are then constructed in section~\ref{sec: BHDW} and we conclude by discussing some possible future directions and include further technical details in two appendices.


\section{Gauged STU Supergravity}	\label{sec: STUsugra}

We consider $D=4$ theories that arise as a truncation of $\mathcal{N}=8$ gauged supergravity \cite{deWit:2002vt,deWit:2007kvg} by requiring invariance under the maximal torus of the relevant gauge group -- see appendix~\ref{app: Cartans} for the expression of the generators of these Cartan subalgebras in terms of the generators of E$_{7(7)}$. 
The bosonic field content of these theories consists of the metric, four vectors and six (pseudo)scalars corresponding to the scalar manifold
\begin{equation}	\label{eq: scalarmanif}
	\left(\frac{\SL(2,\mathbb{R})}{\SO(2)}\right)^3\subset \frac{\rm E_{7(7)}}{\rm SU(8)}\,,
\end{equation}
parametrised by $u_i=\chi_i-ie^{-\varphi_i}$, with $i=1,2,3$.
The bosonic sector of the Lagrangians of these STU supergravities can then be written as
\begin{equation}	\label{eq: STULagrangian}
	\mathcal{L}=(R-V)\,\vol_4+\mathcal{L}_{\rm NLSM}+\mathcal{L}_{\rm vec}\,.
\end{equation}
The scalar kinetic terms read
\begin{equation}	\label{eq: LagrangianNLSM}
	\mathcal{L}_{\rm NLSM}=\tfrac12\sum_i\big[d\varphi_i\wedge*d\varphi_i+e^{2\varphi_i}d\chi_i\wedge*d\chi_i\big]\,,
\end{equation}
and the vector kinetic terms are given by
\begin{equation}
	\mathcal{L}_{\rm vec}=\tfrac12\mathcal{I}_{{\rm ab}} F^{\rm a}\wedge* F^{\rm b}+\tfrac12\mathcal{R}_{{\rm ab}} F^{\rm a}\wedge F^{\rm b}\,,
\end{equation}
with ${\rm a}=1,2,3,4$. These non-minimal couplings can be extracted from the symmetric coset representative of the maximal theory in \eqref{eq: NLSMwithM} via the block decomposition in \eqref{eq: RIinM}~after the identifications in \eqref{eq: TensorHierarchyA}. The result can be given as the period matrix \parbox[t]{3cm}{$\mathcal{N}_{{\rm ab}}=\mathcal{R}_{{\rm ab}}+i\mathcal{I}_{{\rm ab}}$}\footnote{\label{ftn: redefinitions}This period matrix recovers the kinetic and Chern-Simons terms in \cite{Azizi:2016noi} under $\chi_i^{\rm there}=-\chi_i^{\rm here}$ and the following redefinition of the vector fields:
\begin{equation}
		F_1^{\rm here}=F_4^{\rm there}\,,	\qquad
		F_2^{\rm here}=F_3^{\rm there}\,,	\qquad
		F_3^{\rm here}=-F_1^{\rm there}\,,	\qquad
		F_4^{\rm here}=F_2^{\rm there}\,.
\end{equation}
In section \ref{sec: SO8inS7}, this same relabelling applies to the coordinates $\mu^{\rm a}$, $\phi^{\rm a}$, and similarly to $W_{\rm a}$ and $Z_{\rm a}$.}
\begin{equation}
	\mathcal{N}_{{\rm ab}}=\frac{i}W
	\begin{pmatrix}
		-Y^2_1\,Y^2_2\,\tilde{Y}^2_3	&	q_1\,Y^2_1				&	q_3\,\tilde{Y}^2_3					&	q_2\,Y^2_2	\\
		q_1\,Y^2_1				&	-Y^2_1\,\tilde{Y}^2_2\,Y^2_3	&	q_2\,\tilde{Y}^2_2					&	q_3\,Y^2_3	\\
		q_3\,\tilde{Y}^2_3			&	q_2\,\tilde{Y}^2_2			&	-\tilde{Y}^2_1\,\tilde{Y}^2_2\,\tilde{Y}^2_3	&	q_1\,\tilde{Y}^2_1	\\
		q_2\,Y^2_2				&	q_3\,Y^2_3				&	q_1\,\tilde{Y}^2_1					&	-\tilde{Y}^2_1\,Y^2_2\,Y^2_3	\\
	\end{pmatrix}\,,
\end{equation}
in terms of the shorthands \cite{Azizi:2016noi}
\begin{equation}
	\tilde{Y}^2_i=e^{-\varphi_i}+e^{\varphi_i}\,\chi_i^2\,,		\qquad\qquad
	Y^2_i=e^{\varphi_i}\,,		\qquad\qquad
	b_i=e^{\varphi_i}\chi_i\,,					
\end{equation}
and
\begin{equation}	\label{eq: shorthands}
	W=P_0-i\,\tilde{P}_0\,,		\qquad\qquad
	q_i=i\,b_i+b_j\,b_k\,, 	\quad{\rm with}\quad i\neq j\neq k\,,
\end{equation}
with
\begin{equation}
	P_0=1+b_1^2+b_2^2+b_3^2\,,	\qquad\text{and}\qquad
	\tilde{P}_0=2b_1b_2b_3\,.
\end{equation}

For the potential, we consider the parent $\mathcal{N}=8$ supergravity to have one of the following gauge groups: $\SO(8)$ \cite{deWit:1982bul}, its CSO contraction $\SO(6)\ltimes\mathbb{R}^{12}$ \cite{Hull:1984qz,Cordaro:1998tx}  or the dyonic CSO $[\SO(6)\times\SO(2)]\ltimes\mathbb{R}^{12}$ \cite{DallAgata:2011aa,DallAgata:2014tph} gaugings, which all admit a higher-dimensional interpretation. These gaugings can be described by embedding tensors with non-vanishing components in the $\bm{36'}\oplus\bm{36}$ of $\SL(8,\mathbb{R})$ as in \eqref{eq: embtensor912}, with the components $\theta_{AB}$ and $\xi^{AB}$ given by
\begin{equation}	\label{eq: gaugingsClass}
	\begin{aligned}
		\theta&=g\, {\rm diag}(1,1,1,1,1,1,x,x)\,,	\\[5pt]
		\xi&=m\, {\rm diag}(0,0,0,0,0,0,\tilde{x},\tilde{x})\,,\\
	\end{aligned}
\end{equation}
where $g$ (resp. $m$) is the electric (resp. magnetic) coupling constant. By construction, these are valid gaugings satisfying the linear and quadratic constraints for the embedding tensor \cite{deWit:2002vt,Inverso:2016eet} whenever $x \tilde{x}=0$. The three inequivalent choices are
\begin{equation}	\label{eq: gaugings}
	\begin{alignedat}{2}
		\theta_{\8}&=g\, {\rm diag}(1,1,1,1,1,1,1,1)\,,	\qquad \xi_{\8}&&=0\,,\\
		\theta_{\6}&=g\, {\rm diag}(1,1,1,1,1,1,0,0)\,,	\qquad \xi_{\6}&&=0\,,\\
		\theta_{\sst{(6c)}}&=g\, {\rm diag}(1,1,1,1,1,1,0,0)\,,	\qquad \xi_{\sst{(6c)}}&&=m\, {\rm diag}(0,0,0,0,0,0,1,1)\,,
	\end{alignedat}
\end{equation}
with the labels respectively denoting the three different gauge groups above, with $c=m/g\neq0$ in the latter case. 

In the STU truncation, these embedding tensors induce Fayet-Iliopoulos gaugings, whose potentials read
\begin{subequations}	\label{eq: potentials}
	\begin{align}
		V_{\8}&=-4g^2\,\sum_{i}\big(\tilde{Y}^2_i+Y^2_i\big)\,,						\label{eq: potso8}\\[5pt]
		V_{\6}&=V_{\sst{(6c)}}=-4g^2\,\big(\tilde{Y}^2_1+Y^2_2+Y^2_3\big)\,.				\label{eq: potso6so2}
	\end{align}
\end{subequations}
The potential \eqref{eq: potso8} only admits one critical point. It sits at the scalar origin and corresponds to the SO(8) maximally supersymmetric solution. In turn, the potentials \eqref{eq: potso6so2} do not possess any extremum in this sector.

Even though the $V_{\sst{(6c)}}$ potential is blind to the value of the magnetic coupling $m$, the fermion couplings in this theory do depend on it. 
A similar situation has been previously encountered \cite{Lu:2014fpa} in the STU truncation of the dyonically-gauged ${\rm SO}(8)$ supergravity.
In fact, the truncated theory is not supersymmetric for non-vanishing $m$. On the other hand, the electric cases are gauged $\mathcal{N}=2$ supegravities coupled to three vector multiplets. To see this, observe that theories with $\mathcal{N}=2$ supersymmetries can necessarily be recovered in the canonical perspective of \cite{Trigiante:2016mnt, Andrianopoli:1996cm} in terms of special K\"ahler and quaternionic structures. For the scalar manifold \eqref{eq: scalarmanif}, the special holomorphic section can be taken as
\begin{equation}
	\Omega^M(z)=\{1,u_1,u_2,u_3,-u_1u_2u_3,u_2u_3,u_1u_3,u_1u_2\}\,,
\end{equation}
in terms of the special holomorphic coordinates in \eqref{eq: scalarmanif}. This section describes the geometry of the scalar manifold and encodes its K\"ahler potential as
\begin{equation}
	\mathcal{K}=-{\rm log}\big[i\,\bar{\Omega}^M\mathbb{C}_{MN}{\Omega}^N\big]
	=-{\rm log}\big[-i(z_1-\bar{z}_1)(z_2-\bar{z}_2)(z_3-\bar{z}_3)\big]\,,
\end{equation}
with $\mathbb{C}_{MN}$ the symplectic form on ${\rm Sp(8,\mathbb{R})}$.
Purely Fayet-Iliopoulos gaugings have a potential \cite{Trigiante:2016mnt}
\begin{equation}
	V=-4g^2\Big(g^{i\bar i}\partial_iV^M\partial_{\bar i}\bar{V}^N-3V^M\bar{V}^N\Big)\vartheta_M\vartheta_N\,,
\end{equation}
with $V^M=e^{-\mathcal{K}/2}\Omega^M(z)$ a section of the special U(1)-bundle, $g_{i\bar i}=\partial_{i}\partial_{\bar i}\mathcal{K}$ the hermitean metric associated to the K\"ahler potential, and  $\vartheta_M$ the embedding tensor describing how the ${\rm U(1)}$ gauge group sits into the ${\rm SU}(2)$ R-symmetry group. The potential obtained from the truncation of the SO(8) gauging  is given in this language by
\begin{equation}
	\vartheta^{\8}_M=g\,\big(-1,0,0,0,0,1,1,1\big)\,,
\end{equation}
whereas the embedding tensor corresponding to $V_{\6}$ is given by
\begin{equation}
	\vartheta^{\sst{(6)}}_M=g\,\big(-1,0,0,0,0,0,1,1\big)\,.
\end{equation}
For the electric theories, the $\mathcal{N}=2$ supersymmetry variations and fermionic mass-like terms in the Lagrangian associated to these embedding tensors agree with the truncation of the $\mathcal{N}=8$ fermion shifts associated to \eqref{eq: gaugings}. However, for the dyonic gauging in \eqref{eq: gaugings}, the fermion shifts carry dependences on $m$ which can not be recovered in the $\mathcal{N}=2$ language. 

\vspace{8pt}
The potentials in \eqref{eq: potso6so2} can be obtained from $V_{\8}$ by means of a singular scaling. For that, the scalars must be redefined as
\begin{equation}	\label{eq: scalings}
	\begin{alignedat}{2}
		\varphi_1&\mapsto\varphi_1-k	\,,	\qquad	&&\varphi_{2,3}\mapsto\varphi_{2,3}+k\,,	\\[5pt]
		\chi_1&\mapsto e^k\chi_1		\,,	\qquad	&&\chi_{2,3}\mapsto e^{-k}\chi_{2,3}	\,,
	\end{alignedat}
\end{equation}
the gauge coupling as $g\mapsto e^{-k/2}g$, and the gauge fields must also be scaled as
\begin{equation}	\label{eq: scalingvecs}
	A_{1,2,3}\mapsto e^{k/2}A_{1,2,3}\,,	\qquad
	A_4\mapsto e^{-3k/2}A_4\,,
\end{equation}
whilst the metric remains invariant. 
The singular limit $k\to\infty$ on \eqref{eq: STULagrangian} after these redefinitions maps the Lagrangian $\mathcal{L}_\8$ into $\mathcal{L}_\6$, also including the fermion couplings.

\subsection{Tensor and duality hierarchies}

The equations of motion of gauged $D=4$, $\mathcal{N}=8$ supergravity can be written in a formally covariant E$_{7(7)}$--covariant formulation \cite{deWit:2005ub,deWit:2008ta,deWit:2008gc} if one introduces a set of redundant fields in the so-called tensor hierarchy. For generic gaugings of $D=4$, $\mathcal{N}=8$ supergravity, one requires \cite{deWit:2008ta,deWit:2008gc} one-forms in the $\bm{56}$ representation of E$_{7(7)}$, two-forms in the $\bm{133}$ and three-forms in the $\bm{912}$, together with a set of four-forms that will not play a r\^ole in the following. These redundancies can be eliminated at the level of the field strengths through a chain of dualities that relate them to the original fields and their derivatives \cite{Bergshoeff:2009ph,Guarino:2015qaa}, so that combinations of the equations of motion for the original fields are recovered from the Bianchi identities for the forms in the tensor hierarchy. In the following, we show how these reformulations apply to the STU supergravities of interest.

The equations of motion stemming from \eqref{eq: STULagrangian} for the one-forms and scalars are
\begin{equation}	\label{eq: STUEoMs}
	\begin{aligned}
		d\big[\mathcal{I}_{\rm ab}*F^{\rm b}+\mathcal{R}_{\rm ab}F^{\rm b}\big]&=0\,,	\\[5pt]
		d*d\varphi_i-e^{2\varphi_i}d\chi_i\wedge*d\chi_i
		-\tfrac12\partial_{\varphi_i}\mathcal{I}_{{\rm ab}}F^{\rm a}\wedge*F^{\rm b}
		-\tfrac12\partial_{\varphi_i}\mathcal{R}_{\rm ab}F^{\rm a}\wedge F^{\rm b}
		+\partial_{\varphi_i}V\vol_4&=0\,,	\\[5pt]
		d(e^{2\varphi_i}*d\chi_i)
		-\tfrac12\partial_{\chi_i}\mathcal{I}_{\rm ab}F^{\rm a}\wedge*F^{\rm b}
		-\tfrac12\partial_{\chi_i}\mathcal{R}_{\rm ab}F^{\rm a}\wedge F^{\rm b}+\partial_{\chi_i}V\vol_4&=0\,,	
	\end{aligned}
\end{equation}
which can be interpreted as Bianchi identities for two- and three-form field strengths, respectively. For gaugings inside ${\rm SL}(8,\mathbb{R})$, as the ones considered in this work, the $p$-forms in the E$_{7(7)}$ tensor hierarchy are conveniently decomposed into one-forms $A^{AB}$, $\tilde{A}_{AB}$, two-forms $B_A{}^B$, $B_{ABCD}$, and three-forms $C^{AB}$, $\tilde{C}_{AB}$, $C_A{}^{BCD}$, $\tilde{C}^A{}_{BCD}$ in the $\bm{28}\oplus\bm{28'}$, $\bm{63}\oplus\bm{70}$ and $\bm{36}\oplus\bm{36'}\oplus\bm{420}\oplus\bm{420'}$ of ${\rm SL}(8,\mathbb{R})$. Their associated field strengths are given by
\begin{align}	\label{eq: sl8TH}
	H_\2^{AB}&=dA^{AB}+ \theta_{CD}A^{C[A}\wedge A^{B]D}-\xi^{[A\vert C}\tilde{A}_{CD}\wedge A^{D\vert B]}+2\xi^{C[A}B_C{}^{B]}\,,		\nonumber\\[8pt]
	H_{\2AB}&=d\tilde{A}_{AB}-\xi^{CD}\tilde{A}_{C[A}\wedge \tilde{A}_{B]D}+\theta_{[A\vert C}A^{CD}\wedge \tilde{A}_{D\vert B]}-2B_{[A}{}^C\theta_{B]C}\,,	\nonumber\\[8pt]
	H_{\3A}{}^B&=DB_A{}^B+\tfrac12\tilde{A}_{AC}\wedge dA^{CB}+\tfrac12A^{BC}\wedge d\tilde{A}_{CA}		\nonumber\\[3pt]
	&\quad+\tfrac12\theta_{DE}\tilde{A}_{AC}\wedge A^{CD}\wedge A^{EB}+\tfrac{1}{6}\theta_{AC}A^{CD}\wedge \tilde{A}_{DE}\wedge A^{EB}	\nonumber\\[3pt]	
	&\quad-\tfrac12\xi^{CD}\tilde{A}_{AC}\wedge \tilde{A}_{DE}\wedge A^{EB}-\tfrac{1}{6}\xi^{EB}\tilde{A}_{AC}\wedge A^{CD}\wedge\tilde{A}_{DE}	\nonumber\\[3pt]	
	&\qquad+2\,\theta_{AC} C^{CB}-2\,\tilde{C}_{AC} \xi^{CB}-\tfrac{1}{8} \delta_A{}^B({\rm Trace})\,,									\nonumber\\[8pt]	
	H_{\3ABCD}&=DB_{ABCD}+\tfrac14\tilde{A}_{[AB}\wedge d\tilde{A}_{CD]}-\tfrac1{96}\epsilon_{ABCDEFGH}A^{EF}\wedge dA^{GH}		\nonumber\\[3pt]
	&\quad+\tfrac16\Big(\theta_{[A\vert E}A^{EF}-\tilde{A}_{[A\vert E}\xi^{EF}\Big)\wedge\tilde{A}_{F\vert B}\wedge \tilde{A}_{CD]}				\nonumber\\[3pt]
	&\quad+\tfrac1{144}\epsilon_{ABCDEFGH}A^{EF}\wedge\Big(\xi^{GI}\tilde{A}_{IJ}-A^{GI}\theta_{IJ}\Big)\wedge A^{JH}					\nonumber\\[3pt]
	&\quad+\tfrac1{3}\theta_{E[A}\tilde{C}^E{}_{BCD]}-\tfrac1{12}\epsilon_{ABCDEFGH}\xi^{HI}C_I{}^{EFG}\,,
\end{align}
and similarly for the four-forms $H_\4^{AB}$, $\tilde{H}_{\4AB}$, $H_{\4A}{}^{BCD}$, $\tilde{H}_\4^{A}{}_{BCD}$. Here the covariant derivatives are given by the ${\rm SL}(8,\mathbb{R})$ decomposition of
\begin{equation}
	D=d+\Theta_M{}^\alpha A^M\wedge (t_\alpha)_{({\rm R})}\,,
\end{equation}
with $(t_\alpha)_{({\rm R})}$ the E$_{7(7)}$ generators in the appropriate representation. For instance,
\begin{equation}
	\begin{aligned}
		DB_A{}^B&=dB_A{}^B+\theta_{AC}A^{CD}\wedge B_D{}^B+\theta_{CD}A^{BC}\wedge B_A{}^D	\\[5pt]
		&\qquad\qquad+\xi^{CD}\tilde{A}_{CA}\wedge B_D{}^B+\xi^{BD}\tilde{A}_{CD}\wedge B_A{}^C\,,			\\[5pt]
		DB_{ABCD}&=dB_{ABCD}+4B_{[ABC\vert E}\wedge\Big(\xi^{EF}\tilde{A}_{F\vert D]}-A^{EF}\theta_{F\vert D]}\Big)\,.
	\end{aligned}
\end{equation}
 These field strengths obey the Bianchi identities
\begin{align}	\label{eq: sl8Bianchi}
	DH_\2^{AB}&=2\xi^{C[A}H_{\3 C}{}^{B]}\,,		\qquad\qquad
	DH_{\2AB}=-2H_{\3[A}{}^C\theta_{B]C}\,,		\nonumber\\[8pt]
	DH_{\3A}{}^B&=\tilde{H}_{\2 AC}\wedge H_\2^{CB}+2\theta_{AC}H_\4^{CB}-2\tilde{H}_{\4AC}\xi^{CB}-\tfrac18	\delta_A{}^B({\rm Trace})	\,,									\nonumber\\[8pt]	
	DH_{\3ABCD}	&=\tfrac14\tilde{H}_{\2[AB}\wedge \tilde{H}_{\2 CD]}-\tfrac1{96}\epsilon_{ABCDEFGH}H_{\2}^{EF}\wedge H_{\2}^{GH}		\nonumber\\[3pt]
				&\qquad+\tfrac1{3}\theta_{E[A}\tilde{H}_\4^E{}_{BCD]}-\tfrac1{12}\epsilon_{ABCDEFGH}\xi^{HI}H_{\4 I}{}^{EFG}\,,	
\end{align}
which, together with the E$_{7(7)}$ duality relations \cite{Bergshoeff:2009ph,Guarino:2015qaa}
\begin{equation}	\label{eq: E7DH}
	\begin{aligned}
		H_{\2 AB}&=\tfrac12\mathcal{R}_{AB\,CD}\,H_{\2}{}^{CD}+\tfrac12\mathcal{I}_{AB\,CD}\,*H_{\2}{}^{CD}\,,	\\[5pt]
		H_{\3 \alpha}&=-\tfrac1{12}(t_\alpha)_M{}^P\,M_{NP}*DM^{MN}\,,	\\[5pt]
		H_{\4 \alpha}{}^M&=-\tfrac1{84}\Big[(t_\alpha)_P{}^R\,X_{NQ}{}^S\,M^{MN}\Big(M^{PQ}M_{RS}+7\delta^P_S\delta^Q_R\Big)\Big]\Big\vert_{\bm{912}}\vol_4\,,
	\end{aligned}
\end{equation}
recover the equations of motion for the vectors and scalars. Further details on the E$_{7(7)}$ generators $(t_\alpha)_M{}^N$, structure constants $X_{MN}{}^P$ and scalar representative $M_{MN}$ can be found in appendix~\ref{app: Cartans}.

\vspace{5pt}

Demanding invariance under the $\mathcal{H}$ algebra in \eqref{eq: T1234} allows us to consistently truncate this field content to
\begin{equation}	\label{eq: THcontentFull}
	\begin{tabular}{rcl}
		4+4 one-forms				& :	&	$A^{\rm a}\,,\ \tilde{A}_{\rm a}\,,$ 		\\[5pt]
		3+4+6 two-forms			& :	&	$B_p\,,\ B'_{\rm a}\,,\ B_{\rm ab}\,,$				\\[5pt]
		4+4+12+12 three-forms		& :	&	$C_{\rm a}\,,$ $\tilde{C}_{\rm a}\,,$ $C_{\rm ab}\,,$ $\tilde{C}_{\rm ab}\,,$
	\end{tabular}
\end{equation}
with ${\rm a}=1,2,3,4\,,$ and $p=1,2,3$. The forms $B_{\rm ab}$, $C_{\rm ab}$, $\tilde{C}_{\rm ab}$ are off-diagonal, and $B_{\rm ab}$ are also symmetric under exchange of indices. See \eqref{eq: TensorHierarchyA}--\eqref{eq: TensorHierarchyC420} for the relation between the preserved $p$-forms in \eqref{eq: THcontentFull} and the ${\rm SL}(8,\mathbb{R})$ objects.
For convenience, we also introduce an extra two-form $B_4$, constrained as 
\begin{equation}
	B_1+B_2+B_3+B_4=0\,,
\end{equation}
and extend the index as ${\rm a}=(p,4)$.
If we restrict our attention to the class of gaugings in \eqref{eq: gaugingsClass}, we can further truncate consistently the field content to
\begin{equation}	\label{eq: THcontent}
	\begin{tabular}{rcl}
		4+4 one-forms				& :	&	$A^{\rm a}\,,\ \tilde{A}_{\rm a}\,,$ 		\\[5pt]
		3+6 two-forms				& :	&	$B_p\,,\ B_{\rm ab}\,,$					\\[5pt]
		4+4+3+6 three-forms		& :	&	$C_{\rm a}\,,$ $\tilde{C}_{\rm a}\,,$ $C_{p4}\,,$ $\tilde{C}_{p4}\,,$ $\tilde{C}_{12}=\tilde{C}_{43}\,,$ $\tilde{C}_{23}=\tilde{C}_{41}\,,$ $\tilde{C}_{31}=\tilde{C}_{42}\,,$
	\end{tabular}
\end{equation}
with $\tilde{C}_{pq}=\tilde{C}_{qp}$. The field strengths for these gauge potentials can be obtained by implementing these truncations at the level of the ${\rm SL}(8,\mathbb{R})$ field strengths in \eqref{eq: sl8TH}. 
The resulting expressions read
\begin{equation}	\label{eq: THFieldStrengths}
	\begin{aligned}
		H_{\2}^{\rm a}&=dA^{\rm a}\,,	\qquad\qquad
		\tilde{H}_{\2 {\rm a}}=d\tilde{A}_{\rm a}\,,		\\[5pt]
		H_{\3 {\rm a}}&=dB_{\rm a}+2\,\theta_{\rm a}\,C_{\rm a}-2\,\xi_{\rm a}\,\tilde{C}_{\rm a}-\tfrac12A^{\rm a}\wedge d\tilde{A}_{\rm a}-{\tfrac12}\tilde{A}_{\rm a}\wedge dA^{\rm a}	\\[5pt]
		&\qquad\qquad-\tfrac14\sum_{\rm b}\Big[2\,\theta_{\rm b}\,C_{\rm b}-2\,\xi_{\rm b}\,\tilde{C}_{\rm b}-{\tfrac12}A^{\rm b}\wedge d\tilde{A}_{\rm b}-{\tfrac12}\tilde{A}_{\rm b}\wedge dA^{\rm b}\Big]\,,		\\[-2pt]
		H_{\3 {\rm ab}}&=dB_{\rm ab}+\tfrac1{12}\tilde{A}_{\rm (a}\wedge d\tilde{A}_{\rm b)}+\tfrac1{6}\big(\theta_{\rm a}\tilde{C}_{\rm ab}+\theta_{\rm b}\tilde{C}_{\rm ba}\big)	\\[5pt]
		&\qquad\qquad+\tfrac1{24}\sum_{\rm cd}S_{\rm abcd}\big[-A^{\rm c}\wedge dA^{\rm d}+4\,\xi^{\rm c}C_{\rm cd}\big]\,,		\\[-2pt]
		H_{\4 {\rm a}}&=dC_{\rm a}\,,	\qquad\qquad
		\tilde{H}_{\4 {\rm a}}=d\tilde{C}_{\rm a}\,,		
	\end{aligned}
\end{equation}
and similarly for $H_{\4 {\rm ab}}$ and $\tilde{H}_{\4 {\rm ab}}$. In \eqref{eq: THFieldStrengths} there are no sums unless explicitly indicated, and the embedding tensor components $\theta_{\rm a}$ and $\xi_{\rm a}$ are related to $\theta_{AB}$ and $\xi^{AB}$ in \eqref{eq: gaugings} following the same pattern as the three-forms in \eqref{eq: TensorHierarchyC36}. We have also introduced the totally symmetric tensor
\begin{equation}
	S_{\rm abcd}=S_{\rm (abcd)}\,,	\qquad S_{1234}=1\,.
\end{equation}
 From their definition in terms of potentials, it is easy to check that the field strengths satisfy the Bianchi identities
\begin{equation}	\label{eq: THBianchi}
	\begin{aligned}
		dH_{\2}^{\rm a}&=d\tilde{H}_{\2 {\rm a}}=dH_{\4 {\rm a}}=d\tilde{H}_{\4 {\rm a}}=dH_{\4 {\rm ab}}=d\tilde{H}_{\4 {\rm ab}}=0\,,	\\[5pt]
		dH_{\3 {\rm a}}&=2\,\theta_{\rm a}H_{\4 {\rm a}}-2\,\xi_{\rm a}\tilde{H}_{\4 {\rm a}}-H_{\2}^{\rm a}\wedge\tilde{H}_{\2 {\rm a}}-\tfrac14({\rm sum})\,,	\\[5pt]
		dH_{\3 {\rm ab}}&=\tfrac1{12}\tilde{H}_{\rm a}\wedge \tilde{H}_{\rm b}+\tfrac1{6}\big(\theta_{\rm a}\tilde{H}_{\4{\rm ab}}+\theta_{\rm b}\tilde{H}_{\4{\rm ba}}\big)	\\[5pt]
		&\qquad\qquad+\tfrac1{24}\sum_{\rm cd}S_{\rm abcd}\big[-H^{\rm c}\wedge H^{\rm d}+4\,\xi^{\rm c}H_{\4{\rm cd}}\big]\,.
	\end{aligned}
\end{equation}

Implementing the restrictions in \eqref{eq: TensorHierarchyA}--\eqref{eq: TensorHierarchyC420} on \eqref{eq: E7DH}, we can express the dual field strengths in terms of the original STU fields and their derivatives. The magnetic two-form field strengths are then given by
\begin{align}	\label{eq: duality2form}
		\tilde{H}_{\21}&=\frac{1}{\vert W\vert^2}\Big[c_1(\tilde{P}_0-P_0*) H_\2^{1}+Y^2_1(a_1*-\tilde{a}_1) H_\2^{2}+\tilde{Y}^2_3(a_3*-\tilde{a}_3) H_\2^{3}+Y^2_2(a_2*-\tilde{a}_2) H_\2^{4}\Big]\,,	\nonumber\\[5pt]
		\tilde{H}_{\22}&=\frac{1}{\vert W\vert^2}\Big[Y^2_1(a_1*-\tilde{a}_1) H_\2^{1}+c_2(\tilde{P}_0-P_0*) H_\2^{2}+\tilde{Y}^2_2(a_2*-\tilde{a}_2) H_\2^{3}+Y^2_3(a_3*-\tilde{a}_3) H_\2^{4}\Big]\,,	\nonumber\\[5pt]
		\tilde{H}_{\23}&=\frac{1}{\vert W\vert^2}\Big[\tilde{Y}^2_3(a_3*-\tilde{a}_3) H_\2^{1}+\tilde{Y}^2_2(a_2*-\tilde{a}_2) H_\2^{2}+c_3(\tilde{P}_0-P_0*) H_\2^{3}+\tilde{Y}^2_1(a_1*-\tilde{a}_1) H_\2^{4}\Big]\,,	\nonumber\\[5pt]
		\tilde{H}_{\24}&=\frac{1}{\vert W\vert^2}\Big[Y^2_2(a_2*-\tilde{a}_2) H_\2^{1}+Y^2_3(a_3*-\tilde{a}_3) H_\2^{2}+\tilde{Y}^2_1(a_1*-\tilde{a}_1) H_\2^{3}+c_4(\tilde{P}_0-P_0*) H_\2^{4}\Big]\,,
\end{align}
with $W$ in \eqref{eq: shorthands} and
\begin{equation}
	\begin{aligned}
		a_i&=b_{j}b_{k}\big[P_0-2\,b_{i}^2\big]\,,	\qquad\quad
		\tilde{a}_i=b_{i}\,P_0+\frac{\tilde{P}_0^2}{2b_i}\,,	\qquad\text{with}\quad i\neq j\neq k\,,	\\[5pt]
		c_1&=Y_1^2Y_2^2\tilde{Y}_3^2\,,	\qquad
		c_2=Y_1^2\tilde{Y}_2^2Y_3^2\,,	\qquad 
		c_3=\tilde{Y}_1^2\tilde{Y}_2^2\tilde{Y}_3^2\,,	\qquad 
		c_4=\tilde{Y}_1^2Y_2^2Y_3^2\,.
	\end{aligned}
\end{equation}
Similarly, the three-form field strengths are
\begin{equation}	\label{eq: duality3form}
	\begin{aligned}
		&H_{\31}=\frac{1}{2}*(-d\varphi_1- d\varphi_2+d\varphi_3+e^{2\varphi_1}\chi_1 d\chi_1+e^{2\varphi_2}\chi_2 d\chi_2-e^{2\varphi_3}\chi_3 d\chi_3)\,,	\\[5pt]
		&H_{\32}=\frac{1}{2}*(-d\varphi_1+ d\varphi_2-d\varphi_3+e^{2\varphi_1}\chi_1 d\chi_1-e^{2\varphi_2}\chi_2 d\chi_2+e^{2\varphi_3}\chi_3 d\chi_3)\,,	\\[5pt]
		&H_{\33}=\frac{1}{2}*(d\varphi_1+ d\varphi_2+d\varphi_3-e^{2\varphi_1}\chi_1 d\chi_1-e^{2\varphi_2}\chi_2 d\chi_2-e^{2\varphi_3}\chi_3 d\chi_3)\,,	\\[5pt]
		&H_{\34}=\frac{1}{2}*(d\varphi_1- d\varphi_2-d\varphi_3-e^{2\varphi_1}\chi_1 d\chi_1+e^{2\varphi_2}\chi_2 d\chi_2-e^{2\varphi_3}\chi_3 d\chi_3)\,.
	\end{aligned}
\end{equation}
for the ones in the $\bm{63}$ of ${\rm SL}(8,\mathbb{R})$, and
\begin{equation}	\label{eq: duality3form70}
	\begin{gathered}
		H_{\312}=\frac{1}{12}*e^{2\varphi_1} d\chi_1\,,	\qquad\quad
		H_{\314}=\frac{1}{12}*e^{2\varphi_2} d\chi_2\,,	\qquad\quad
		H_{\324}=\frac{1}{12}*e^{2\varphi_3} d\chi_3\,,	\\[5pt]
		H_{\334}=\frac{1}{12}*\big[2\chi_1d\varphi_1+(1-e^{-2\varphi_1}\chi_1^2) d\chi_1\big]\,,	\\[5pt]
		H_{\323}=\frac{1}{12}*\big[2\chi_2d\varphi_2+(1-e^{-2\varphi_2}\chi_2^2) d\chi_2\big]\,,	\\[5pt]
		H_{\313}=\frac{1}{12}*\big[2\chi_3d\varphi_3+(1-e^{-2\varphi_3}\chi_3^2) d\chi_3\big]\,,	\\[-14pt]
	\end{gathered}
\end{equation}\\[-10pt]
for the ones in the $\bm{70}$.
The expressions for the four-form field strengths in terms of the scalars depend on the choice of embedding tensor. Employing \eqref{eq: E7DH} for the class of gaugings in \eqref{eq: gaugingsClass}, we obtain
\begin{equation}	\label{eq: duality4form}
	\begin{aligned}
		H_{\41}&=2g\big[\tilde{Y}_1^2+x\,\tilde{Y}_2^2+Y_3^2\big]\vol_4\,,	\qquad\quad
		\tilde{H}_{\41}=2m\,\tilde{x}\,Y_2^2\,\vol_4\,,		\\[5pt]
		H_{\42}&=2g\big[\tilde{Y}_1^2+Y_2^2+x\,\tilde{Y}_3^2\big]\vol_4\,,	\qquad\quad
		\tilde{H}_{\42}=2m\,\tilde{x}\,Y_3^2\,\vol_4\,,		\\[5pt]
		H_{\43}&=2g\big[x\,Y_1^2+Y_2^2+Y_3^2\big]\vol_4\,,	\qquad\quad
		\tilde{H}_{\43}=2m\,\tilde{x}\,\tilde{Y}_1^2\,\vol_4\,,		\\[5pt]
		H_{\44}&=2g\big[Y_1^2+\tilde{Y}_2^2+\tilde{Y}_3^2\big]\vol_4\,,		\hspace{1.43cm}
		\tilde{H}_{\44}=0\,,		
	\end{aligned}
\end{equation}
and
\begin{equation}	\label{eq: duality4form420}
	\begin{alignedat}{3}
		&H_{\434}=2m\tilde{x} \chi_1e^{\varphi_1}\vol_4\,,	\qquad\quad
		&&H_{\414}=2m\tilde{x} \chi_2e^{\varphi_2}\vol_4\,,	\qquad\quad
		&&H_{\424}=2m\tilde{x} \chi_3e^{\varphi_3}\vol_4\,,	\\[5pt]
		&\tilde H_{\412}=-2g \chi_1e^{\varphi_1}\vol_4\,,	\qquad\quad
		&&\tilde H_{\423}=-2g \chi_2e^{\varphi_2}\vol_4\,,	\qquad\quad
		&&\tilde H_{\431}=-2g \chi_3e^{\varphi_3}\vol_4\,,	\\[5pt]
		&\tilde H_{\434}=-2gx \chi_1e^{\varphi_1}\vol_4\,,	\qquad\quad
		&&\tilde H_{\414}=-2gx \chi_2e^{\varphi_2}\vol_4\,,	\qquad\quad
		&&\tilde H_{\424}=-2gx \chi_3e^{\varphi_3}\vol_4\,.
	\end{alignedat}
\end{equation}
The forms in \eqref{eq: duality4form} can be used to reproduce the potentials in \eqref{eq: potentials} as
\begin{equation}
	\sum_{\rm a}\big[\theta_{\rm a}H_{\4\rm a}+\xi^{\rm a}\tilde{H}_{\4\rm a}\big]=-V\vol_4\,.
\end{equation}

As noted before, inserting the duality relations \eqref{eq: duality2form} into the Bianchi identities \eqref{eq: THBianchi}, we obtain the scalar-Maxwell equations in \eqref{eq: STUEoMs} stemming from the Lagrangian \eqref{eq: STULagrangian}. Similarly, the Bianchi identities for the three-form field strengths are identically verified if the equations of motion for the scalars are satisfied. The derivatives of the scalars that appear in equations of motion for the dilatons are retrieved from
\begin{equation}	
	\begin{aligned}
		H_{\31}+H_{\32}=-2\,G_{uv}\,k^u[h_1] *d\Phi^v\,,&	\qquad\quad
		H_{\31}+H_{\34}=-2\,G_{uv}\,k^u[h_2] *d\Phi^v\,,	\\[5pt]
		H_{\32}+H_{\34}&=-2\,G_{uv}\,k^u[h_3] *d\Phi^v\,,
	\end{aligned}
\end{equation}
with $\Phi^u$ a collective name for the six scalars $(\varphi_i,\, \chi_i)$ and $G_{uv}$ the non-linear sigma-model metric in \eqref{eq: LagrangianNLSM}. The Killing vectors $k[h_i]$ are defined in \eqref{eq: NLSMKilling}. Similarly, the three-forms in \eqref{eq: duality3form70} are related to $k[e_i]$ and $k[f_i]$ in \eqref{eq: NLSMKilling} and encode the derivatives of the axions in their equations of motion. 
The derivatives of the scalar potential in \eqref{eq: STUEoMs} are accounted for by the four-form field strengths as
\begin{equation}
	\begin{aligned}
		\theta_1H_{\41}+\theta_2H_{\42}-\xi_1\tilde{H}_{\41}-\xi_2\tilde{H}_{\42}-\tfrac12\sum_{\rm a}\big(\theta_aH_{\4 {\rm a}}-\xi_a\tilde{H}_{\4 {\rm a}}\big)=&\,\tfrac14 k^u[h_1]\partial_uV\, \vol_4\,,	\\[5pt]
		\theta_1H_{\41}+\theta_4H_{\44}-\xi_1\tilde{H}_{\41}-\xi_4\tilde{H}_{\44}-\tfrac12\sum_{\rm a}\big(\theta_aH_{\4 {\rm a}}-\xi_a\tilde{H}_{\4 {\rm a}}\big)=&\,\tfrac14 k^u[h_2]\partial_uV\, \vol_4\,,	\\[5pt]
		\theta_2H_{\42}+\theta_4H_{\44}-\xi_2\tilde{H}_{\42}-\xi_4\tilde{H}_{\44}-\tfrac12\sum_{\rm a}\big(\theta_aH_{\4 {\rm a}}-\xi_a\tilde{H}_{\4 {\rm a}}\big)=&\,\tfrac14 k^u[h_3]\partial_uV\, \vol_4\,,	\\[5pt]
		g\tilde{H}_{\412}=\tfrac1{28}k^u[e_1]\partial_uV\vol_4\,,	\quad
		g\tilde{H}_{\423}=\tfrac1{28}k^u[e_2]\partial_uV\vol_4\,,	\quad
		g\tilde{H}_{\431}&=\tfrac1{28}k^u[e_3]\partial_uV\vol_4\,,
	\end{aligned}
\end{equation}
and, finally, the non-minimal couplings to the vector fields follow from
\begin{equation}
	\begin{aligned}
		H_\2^1\wedge \tilde{H}_{\21}+H_\2^2\wedge \tilde{H}_{\22}-\tfrac12\sum_{\rm a}H_{\2}^{\rm a}\wedge \tilde{H}_{\2 {\rm a}}&=\tfrac14k^u[h_1]\partial_u\Big(\sum_{\rm a}H_{\2}^{\rm a}\wedge \tilde{H}_{\2{\rm a}}\Big)\,,		\\[5pt]
		H_\2^1\wedge \tilde{H}_{\21}+H_\2^4\wedge \tilde{H}_{\24}-\tfrac12\sum_{\rm a}H_{\2}^{\rm a}\wedge \tilde{H}_{\2 {\rm a}}&=\tfrac14k^u[h_2]\partial_u\Big(\sum_{\rm a}H_{\2}^{\rm a}\wedge \tilde{H}_{\2{\rm a}}\Big)\,,	\\[5pt]
		H_\2^2\wedge \tilde{H}_{\22}+H_\2^4\wedge \tilde{H}_{\24}-\tfrac12\sum_{\rm a}H_{\2}^{\rm a}\wedge \tilde{H}_{\2 {\rm a}}&=\tfrac14k^u[h_3]\partial_u\Big(\sum_{\rm a}H_{\2}^{\rm a}\wedge \tilde{H}_{\2{\rm a}}\Big)\,,	
	\end{aligned}
\end{equation}
in the equation for the dilations, and
\begin{equation}
	\begin{aligned}
		\tilde{H}_{\21}\wedge\tilde{H}_{\22}-H_\2^3\wedge H_\2^4&=\tfrac12k^u[e_1]\partial_u\Big(\sum_{\rm a}H_{\2}^{\rm a}\wedge \tilde{H}_{\2{\rm a}}\Big)\,,	\\[5pt]
		\tilde{H}_{\21}\wedge\tilde{H}_{\24}-H_\2^2\wedge H_\2^3&=\tfrac12k^u[e_2]\partial_u\Big(\sum_{\rm a}H_{\2}^{\rm a}\wedge \tilde{H}_{\2{\rm a}}\Big)\,,	\\[5pt]
		\tilde{H}_{\22}\wedge\tilde{H}_{\24}-H_\2^1\wedge H_\2^3&=\tfrac12k^u[e_3]\partial_u\Big(\sum_{\rm a}H_{\2}^{\rm a}\wedge \tilde{H}_{\2{\rm a}}\Big)\,,
	\end{aligned}
\end{equation}
in the one for the axions.

\vspace{8pt}

Turning now our attention to the singular limit relating the different gaugings, under the scaling \eqref{eq: scalings} the three-form field strengths in \eqref{eq: duality3form} stay invariant, the magnetic two-form field strengths have opposite scaling to the electric forms in \eqref{eq: scalingvecs},
\begin{equation}	
	\tilde{H}_{\21,2,3}\mapsto e^{-k/2}\tilde{H}_{\21,2,3}\,,	\qquad
	\tilde{H}_{\24}\mapsto e^{3k/2}\tilde{H}_{\24}\,,
\end{equation}
and the electric four-form field strengths reduce to
\begin{equation}	\label{eq: H4scalings}
	H_{\4 \rm a}\mapsto e^{k}\bar H_{\4 \rm a}\,,
\end{equation}
with $\bar H_{\4 \rm a}$ the four-forms given by \eqref{eq: duality4form} with $x=0$ and $\tilde{x}=0$. The scaling of the other four-forms, which will not be relevant in the following, can be computed in the same way from \eqref{eq: duality4form} and \eqref{eq: duality4form420}.

In the sequel, we will embed the previous $D=4$ gauged supergravities into type IIB and $D=11$ supergravity using a duality-covariant reformulation of the latter higher-dimensional theories known as Exceptional Field Theory, and encoding part of the dependence on the $4d$ fields in terms of a subset of the forms in the tensor hierarchy.


\section{E$_{7(7)}$ Exceptional Field Theory}	\label{sec: E77ExFT}

The bosonic field content of ExFT \cite{Hohm:2013uia} is given by
\begin{equation}	\label{eq: ExFTfields}
	\{\bm{e}_\mu{}^a\,,\ \bm{M}_{MN}\,,\ \bm{A}_\mu{}^M\,,\ \bm{B}_{\mu\nu}{}_\alpha\,,\ \bm{B}_{\mu\nu}{}_M\}\,,
\end{equation}
with all fields depending on both ``external'', $x^\mu$, $\mu=0,\dots,3$, and extended ``internal'' coordinates, $Y^M$, $M=1,\dots,56$. This dependence and the fields themselves are restricted by the section constraints \cite{Hohm:2013uia}
\begin{equation}	\label{eq: E77sc}
	(t_\alpha)^{MN}Q_M\otimes Q_N=0\,,	\qquad
	\Omega^{MN}Q_M\otimes Q_N=0\,,
\end{equation}
with $Q_M\in\{\partial_M\,,\ \bm{B}_{\mu\nu}{}_M\}$ and the derivatives acting on any combination of fields or gauge parameters in the theory. Here, $(t_\alpha)_M{}^N$ are the algebra generators and indices are raised and lowered with the invariant symplectic form $\Omega_{MN}$ as
\begin{equation}	\label{eq: E77omega}
	\Omega^{MN}=\begin{pmatrix}
					0	&	\mathbbm{1}_{28}	\\
					-\mathbbm{1}_{28}	&	0
				\end{pmatrix}\,,		\qquad
	V^M=\Omega^{MN}V_N\,,\qquad
	V_M=V^N\Omega_{NM}\,.
\end{equation}
These constraints are needed for the generalised Lie derivative to close into a local E$_{7(7)}$ gauge algebra. Variations under the latter are given by \cite{Hohm:2013uia}
\begin{equation}	\label{eq: genLie}
	\delta_\Lambda V^M\equiv\mathcal{L}_{\Lambda}V^M=\Lambda^N\partial_NV^M-12\,\mathbb{P}^M{}_N{}^K{}_L\,\partial_K\Lambda^L\, V^N+\lambda(V)\, \partial_N\Lambda^N\,V^M\,,
\end{equation}
where $\mathbb{P}^M{}_N{}^K{}_L=(t_\alpha)_N{}^M\,(t^\alpha)_L{}^K$ is the projector onto the adjoint representation and $\lambda(V)$ is the weight associated to the generalised vector $V^M$. To solve the constraints \eqref{eq: E77sc}, E$_{7(7)}$ can be reduced down to ${\rm GL}(7,\mathbb{R})$ (M-theory section) or ${\rm GL}(6,\mathbb{R})\times{\rm SL}(2,\mathbb{R})$ (Type IIB section).
After this reduction, the variation $\delta_\Lambda$ encodes the behaviour of the different fields under both ``internal'' diffeomorphisms and gauge transformations.

\subsection{M-theory section}

For $D=11$ supergravity, we use conventions in which our fields are subject to the action
\begin{equation}
	S=\int d^{11}x\sqrt{\vert\hat g_{11}\vert}\Big[\hat R_{11}-\tfrac12\vert\hat F_\4\vert^2\Big]-\tfrac16\int \hat A_\3\wedge \hat F_\4\wedge \hat F_\4\,,
\end{equation}
with $\hat F_\4=d\hat A_\3$ and contraction of indices with weight one denoted by $\vert \cdot\vert^2$, i.e.
\begin{equation}	\label{eq: contraction}
	\vert\hat F_{\sst{(p)}}\vert^2=\tfrac1{p!}\hat{F}_{\hat\mu_1\dots\hat\mu_p}\,\hat{F}^{\hat\mu_1\dots\hat\mu_p}\,.
\end{equation}

Under the structure group relevant for a seven-dimensional internal space, the extended ExFT coordinates decompose as\footnote{The coordinate index in $y^i$ should not be confused with the index labelling the different factors in \eqref{eq: scalarmanif}.}
\begin{equation}	\label{eq: E77fundintoGL7}
	\begin{tabular}{ccccc}
		E$_{7(7)}$	&	$\supset$	&	${\rm SL}(8,\mathbb{R})$	&	$\supset$	&	${\rm GL}(7,\mathbb{R})$			\\[5pt]
		$\bm{56}$		&	$\to$		&	$\bm{28}\oplus\bm{28'}$	&	$\to$		&	$\bm{7}_{+3}\oplus\bm{21'}_{+1}\oplus\bm{21}_{-1}\oplus\bm{7'}_{-3}$\;,	\\[5pt]
		$\{Y^M\}$		&	$\to$		&	$\{Y^{AB},\, Y_{AB}\}$	&	$\to$		&	$\{y^i\,,\ y_{ij}\,,\ y^{ij}\,,\ y_{i}\}$\;,
	\end{tabular}
\end{equation}
with $y^i=Y^{i8}$, etc. The section constraint \eqref{eq: E77sc} can be solved by imposing
\begin{equation}	\label{eq: GL7sc}
	\partial^{ij}=\partial_{ij}=\partial^i=0\,, \qquad{\rm and}\qquad
	\bm{B}_{\mu\nu}{}^{ij}=\bm{B}_{\mu\nu}{}_{ij}=\bm{B}_{\mu\nu}{}^i=0\,,
\end{equation}
keeping only the $\bm{B}_{\mu\nu}{}_i$ components of $\bm{B}_{\mu\nu}{}_M$. Similarly, the objects in the adjoint representation of E$_{7(7)}$ break according to
\begin{equation}	\label{eq: E77adjintoGL7}
	\begin{tabular}{ccccc}
		E$_{7(7)}$	&	$\supset$	&	${\rm SL}(8,\mathbb{R})$	&	$\supset$	&	${\rm GL}(7,\mathbb{R})$			\\[5pt]
		$\bm{133}$		&	$\to$		&	$\bm{63}\oplus\bm{70}$	&	$\to$		&	$\bm{7'}_{+4}\oplus\bm{35}_{+2}\oplus(\bm{48}\oplus\bm{1})_{0}\oplus\bm{35'}_{-2}\oplus\bm{7}_{-4}$\;,	\\[5pt]
		$\{t_\alpha\}$		&	$\to$		&	$\{t_A{}^B,\, t_{ABCD}\}$	&	$\to$		&	$\{t_i\,,\ t^{ijk}\,,\ t_0\,,\ t_i{}^j\,,\ t_{ijk}\,,\ t^i\}$\;.
	\end{tabular}
\end{equation}

To make contact with M-theory, one needs to split the $D=11$ structure group ${\rm GL}(11,\mathbb{R})\supset{\rm GL}(4,\mathbb{R})\times{\rm GL}(7,\mathbb{R})$. Then, the metric and three-form of $D=11$ supergravity give rise to the following fields:
\begin{equation}	\label{eq: GL7Mtheory}
	\{g_{\mu\nu}\,,\ A_\mu{}^i\,,\ \phi_{ij}\,;\ C_{ijk}\,,\ C_{ij\rho}\,,\ C_{i\nu\rho}\,,\ C_{ijklmn}\}\,,
\end{equation}
where all of the fields depend on both $x^\mu$ and $y^i$. The forms $C_{ijk}\,,\ C_{ij\rho}\,,\ C_{i\nu\rho}$ are related to the components of the eleven dimensional $\hat{A}_\3$ through the usual Kaluza-Klein decomposition with flattening and unflattening of indices. The $C_{ijklmn}$ components are dual to the external legs $C_{\mu\nu\rho}$ through
\begin{equation}	\label{eq: dualFR}
	\begin{aligned}
		F_{\7}&=*_{11} dC^{\rm \sst{(ext)}}_\3	\\
			&=dC_{\6}-\frac12C_{\3}\wedge dC_{\3}\,,	
	\end{aligned}
\end{equation}
with
\begin{equation}	\label{eq: GL7dictGauge}
	C^{\rm \sst{(ext)}}_{\3}=\tfrac1{3!}C_{\mu\nu\rho}dx^\mu\wedge dx^\nu\wedge dx^\rho\,,	\qquad
	C_{\3}=\tfrac1{3!}C_{ijk}dy^i\wedge dy^j\wedge dy^k\,,
\end{equation}
as required by the infinitesimal transformations
\begin{equation}	\label{eq: GL7dictGauge}
	\delta C_{\3}=L_\xi C_{\3}+d\Lambda_{\2}\,,	\qquad
	\delta C_{\6}=L_\xi C_{\6}+d\Lambda_{\5}-\frac12d\Lambda_{\2}\wedge C_{\3}\,.
\end{equation}

These fields are then related to the ExFT fields branched according to \eqref{eq: E77fundintoGL7} and \eqref{eq: E77adjintoGL7}. For the tensor-like degrees of freedom, the dictionary reads
\begin{equation}	\label{eq: GL7dictTensor}
	\bm{e}_\mu{}^a=\phi^{1/4}e_\mu{}^a\,,	\qquad
	\bm{A}_\mu{}^i=A_\mu{}^i\,,	\qquad
	\bm{A}_\mu{}_{ij}=C_{\mu ij}\,,
\end{equation}
and $C_{\mu\nu i}$ are related to $\bm{B}_{\mu\nu\, \alpha}$ and $\bm{B}_{\mu\nu\, M}$. For the $D=4$ scalars,
\begin{equation}	\label{eq: GL7dictScalar}
	\bm{M}^{ij}=\phi^{-1/2}\phi^{ij}\,,	\quad
	\bm{M}^{i}{}_{jk}=3\phi^{-1/2}\phi^{il}C_{ljk}\,,	\quad
	\bm{M}^{i}{\,}^{jk}=-\frac3{20}\phi^{-1/2}\phi^{i[j}\varepsilon^{k]l_1l_2l_3l_4l_5l_6}C_{l_1l_2l_3l_4l_5l_6}
\end{equation}
for the components of the $\bm{M}^{MN}$ coset representative of E$_{7(7)}$/SU(8). Here and throughout, $\phi={\rm det}\,\phi_{ij}$. Conversely, the internal components of the three- and six-form can be given~as
\begin{equation}	\label{eq: GL7dictScalarII}
	C_{ijk}=\frac13\phi^{1/2}\phi_{l[i}\bm{M}^l{}_{jk]}\,,	\qquad
	C_{l_1l_2l_3l_4l_5l_6}=-\frac{20}3\phi^{1/2}\phi_{ij}\bm{M}^{i\; jk}\varepsilon_{kl_1l_2l_3l_4l_5l_6}\,.
\end{equation}

In the following, we will show that in the context of consistent truncations, one can circumvent the dualisation in \eqref{eq: dualFR} by working in terms of the $p$-forms of the four-dimensional tensor hierarchy, which also cleanly account for the information corresponding to the $4d$ vectors $A_{\mu}^i$ and $C_{\mu ij}$, and two-forms $C_{\mu\nu i}$.

\subsection{Type IIB section}

Similarly to the $D=11$ case, the extended ExFT coordinates decompose under ${\rm GL}(6,\mathbb{R})\times\SL(2,\mathbb{R})$ following
\begin{equation}	\label{eq: E77fundintoGL6xSL2}
	\begin{tabular}{ccc}
		E$_{7(7)}$	&	$\supset$		&	${\rm GL}(6,\mathbb{R})\times\SL(2,\mathbb{R})$ 		\\[5pt]
		$\bm{56}$		&	$\to$			&	$(\bm{6},\bm{1})_{+2}\oplus(\bm{6}',\bm{2})_{+1}\oplus(\bm{20},\bm{1})_{0}\oplus(\bm{6},\bm{2})_{-1}\oplus(\bm{6}',\bm{1})_{-2}$	\;,	\\[5pt]
		$\{Y^M\}$		&	$\to$			&	$\{y^i\,,\ y_{ia}\,,\ y^{ijk}\,,\ y^{ia}\,,\ y_{i}\}$\;,
	\end{tabular}
\end{equation}
now with $i=1,\dots,6$ and $a=1,2$. The section constraint \eqref{eq: E77sc} can be solved by requiring that all fields and parameters only depend on $y^i$ and that the only non-zero component of the constrained two-form is $\bm{B}_{\mu\nu}{}_i$. In turn, the objects in the adjoint representation adhere~to
\begin{equation}	\label{eq: E77adjintoGL6xSL2}
	\begin{tabular}{ccc}
		E$_{7(7)}$	&	$\supset$		&	${\rm GL}(6,\mathbb{R})\times\SL(2,\mathbb{R})$ 		\\[5pt]
		$\bm{133}$	&	$\to$			&	$(\bm{1},\bm{2})_{3}\oplus(\bm{15'},\bm{1})_{2}\oplus(\bm{15},\bm{2})_{1}\oplus(\bm{35}+\bm{1},\bm{1})_{0}\hbox{\hspace{1.5cm}}$	\\[2pt]		
								&&	$\hbox{\hspace{1.5cm}}\oplus(\bm{1},\bm{3})_{0}\oplus(\bm{15'},\bm{2})_{-1}\oplus(\bm{15},\bm{1})_{-2}\oplus(\bm{1},\bm{2})_{-3}$	\;,	\\[5pt]
		$\{t_\alpha\}$		&	$\to$			&	$\{t_a\,,\ t_{ij}\,,\ t^{ij\, a}\,,\ t_{i}{}^j\,,\ t_{0}\,,\ t_{a}{}^b\,,\ t_{ij\, a}\,,\ t^{ij}\,,\ \tilde{t}_{a}\}$\;.
	\end{tabular}
\end{equation}

Contact with type IIB supergravity is achieved after splitting the ten-dimensional structure group ${\rm GL}(10,\mathbb{R})$ into ${\rm GL}(4,\mathbb{R})\times{\rm GL}(6,\mathbb{R})$ so that the bosonic fields read
\begin{equation}	\label{eq: GL6IIB}
	\{g_{\mu\nu}\,,\ \phi_{ij}\,,\ \Phi\,,\ C_{0}\,,\ C_{\mu\nu}{}^a\,,\ C_{\mu i}{}^a\,,\ C_{ij}{}^a\,,\ C_{\mu\nu\rho\sigma}\,,\ C_{\mu\nu\rho i}\,,\ C_{\mu\nu ij}\,,\ C_{\mu ijk}\,,\ C_{ijkl}\}\,,
\end{equation}
again with all of them depending both on $x^\mu$ and $y^i$ and taking into account the flattening and unflattening of indices with the Kaluza-Klein vector. Here, we use conventions in which the type IIB pseudoaction is given by
\begin{equation}	\label{eq: IIBpaction}
	S=\int d^{10}x\sqrt{\vert\hat g_{10}\vert}\Big[\hat R_{10}-\tfrac12(\partial\hat\Phi)^2-\tfrac12e^{2\hat\Phi}\vert\hat F_\1\vert^2
	-\tfrac12e^{-\hat\Phi}\vert \hat H_\3\vert^2-\tfrac12e^{\hat\Phi}\vert\hat F_\3\vert^2-\tfrac1{4}\vert\hat F_\5\vert^2\Big]
	+\int \mathcal{L}_{\rm top}\,,
\end{equation}
with $\vert\hat F_{\sst{(p)}}\vert^2$ and $\vert\hat H_{\3}\vert^2$ defined in \eqref{eq: contraction}, and a topological term $\mathcal{L}_{\rm top}=\hat C_\4\wedge\hat H_\3\wedge\hat F_\3$, and the field strengths given by
\begin{equation}
	\hat H_\3=d\hat B_\2\,,	\quad
	\hat F_\1=d\hat C_\0\,,	\quad
	\hat F_\3=d\hat C_\2-\hat C_\0\, \hat H_\3\,,	\quad
	\hat F_\5=d\hat C_\4+\frac12\epsilon_{ab}\,\hat C_\2^a\wedge d\hat C_\2^b\,,
\end{equation}
with $\hat C_\2^a=(\hat B_\2, \hat C_\2)$. In addition to the equations of motion obtained from \eqref{eq: IIBpaction}, the self-duality of $\hat F_\5$ needs also to be imposed.
The relation to the ExFT fields branched according to \eqref{eq: E77fundintoGL6xSL2} and \eqref{eq: E77adjintoGL6xSL2} then is
\begin{equation}	\label{eq: GL6xSL2dictScalar}
	\begin{aligned}
		\bm{M}^{ij}&=\phi^{-1/2}\phi^{ij}\,,	\qquad\qquad
		\bm{M}^{i}{}_{ja}=\phi^{-1/2}\phi^{ik}\epsilon_{ab}C_{kj}{}^{b}\,,		\\[5pt]
		\bm{M}^{i}{}_{jkl}&=\phi^{-1/2}\phi^{im}\big(24C_{mjkl}+3\epsilon_{ab}C_{m[j}{}^a C_{kl]}{}^b\big)\,,	\\[5pt]
		\bm{M}_{ia}{}_{jb}&=6\,\phi^{-1/2}\phi_{ij}\big(\mathfrak{m}_{ab}-\epsilon_{ac}\epsilon_{bd}C_{kl}{}^c C^{kl}{}^d\big)\,,
	\end{aligned}
\end{equation}
for the $D=4$ scalars, with the IIB axiodilation encoded in the $\SL(2,\mathbb{R})$ matrix as 
\begin{equation}	\label{eq: SL2m}
	\mathfrak{m}=e^{\Phi}\begin{pmatrix} e^{-2\Phi}+C_0^2 & -C_0\\ -C_0&1 \end{pmatrix}\,.
\end{equation}
Similarly, we have $\bm{e}_\mu{}^a=\phi^{1/4}e_\mu{}^a$ and $\bm{A}_\mu{}^i=A_\mu{}^i$ for the tensor-like components of the metric, and the other $p$-form contributions we will phrase in terms of objects in the tensor hierarchy when we turn to consistent truncations in the sequel.

To make contact with the $S^5\times S^1$ configurations, we need to further decompose under ${\rm GL}(5,\mathbb{R})\times{\rm GL}(1,\mathbb{R})\times\SL(2,\mathbb{R})$ via $i=(\i,6)$ with $\i=1,\dots,5$. This subgroup is common to both $\SL(8,\mathbb{R})$ and ${\rm GL}(6,\mathbb{R})\times\SL(2,\mathbb{R})$, and therefore its representations can be given in terms of $\SL(8,\mathbb{R})$ pairs. In our conventions, 
\begin{equation}	\label{eq: GL5embeddings}
	\begin{aligned}
		V^i=(V^{\i6},\ V_{78})\,,	\qquad
		V^{ijk}&=(\epsilon^{ijk\l\m6}V_{\l\m},\ V^{\i\j}\delta^k_6)\,,	\qquad
		V_i=(V_{\i6},\ V^{78})\,,	\\[7pt]
		V_{ia}=&(V_{\i a},\ V^{6 a})\,,	\qquad
		V^{ia}=(V^{\i a},\ V_{6 a})\,,
	\end{aligned}
\end{equation}
and for the coordinates we further introduce $y^{\i}=\hat{y}^{\i}$ and $y^{6}=\tilde{y}$.

\subsection{Generalised Scherk-Schwarz Ans\"atze}

The ExFT fields \eqref{eq: ExFTfields} can be parametrised in terms of $D=4$ $\mathcal{N}=8$ supergravity fields via a Scherck-Schwarz Ansatz \cite{Hohm:2014qga}\footnote{In the following, we add bars to the E$_{7(7)}$ and ${\rm SL}(8,\mathbb{R})$ indices in the previous section to distinguish flat and curved counterparts.}
\begin{equation}	\label{eq: SSAnsatz}
	\begin{aligned}
		\bm{g}_{\mu\nu}(x,Y)&=\rho^{-2}(Y)\,g_{\mu\nu}(x)	\,,	\\[5pt]
		\bm{M}_{MN}(x,Y)&=U_M{}^{\bar M}(Y)U_N{}^{\bar N}(Y)M_{\bar M\bar N}(x)\,,		\\[5pt]
		\bm{A}_\mu{}^M(x,Y)&=\rho^{-1}(Y)(U^{-1})^M{}_{\bar M}(Y)A_\mu{}^{\bar M}(x)\,,		\\[5pt]
		\bm{B}_{\mu\nu}{}_\alpha(x,Y)&=\rho^{-2}(Y)U_\alpha{}^{\bar \alpha}(Y)B_{\mu\nu}{}_{\bar \alpha}(x)	\,,	\\[5pt]
		\bm{B}_{\mu\nu}{}_M(x,Y)&=-2\rho^{-2}(Y)(U^{-1})^P{}_{\bar S}(Y)\partial_MU_P{}^{\bar R}(Y)(t^{\bar\alpha})_{\bar R}{}^{\bar S}B_{\mu\nu}{}_{\bar \alpha}(x)\,.		\\[5pt]
	\end{aligned}
\end{equation}
This Ansatz provides a consistent truncation of the ExFT equations of motion down to $D=4$ maximal supergravity provided that the twist matrix and scaling factor define a generalised frame $\mathcal{U}_{\bar M}{}^M=\rho^{-1}(U^{-1})^M{}_{\bar M}$ such that
\begin{equation}	\label{eq: parallelisation}
	\mathcal{L}_{\mathcal{U}_{\bar M}}\mathcal{U}_{\bar N}=X_{\bar M\bar N}{}^{\bar P}\, \mathcal{U}_{\bar P}\,,
\end{equation}
with $\mathcal{L}$ given in \eqref{eq: genLie} and $X_{\bar M\bar N}{}^{\bar P}$ a set of constants to be identified with the embedding tensor of the lower-dimensional supergravity, specified in our cases by \eqref{eq: gaugings} through $X_{\bar M\bar N}{}^{\bar P}=\Theta_{\bar M}{}^{\bar \alpha}(t_{\bar \alpha})_{\bar N}{}^{\bar P}$, with $\Theta_{\bar M}{}^{\bar \alpha}$ given in \eqref{eq: embtensor912} in terms of the components in \eqref{eq: gaugingsClass}.

For the gaugings under consideration, the twist matrix can be written as
\begin{align}		\label{eq: twist}
	U_M{}^{\bar{M}}(y)
		&=
		\left(
		\begin{array}{c;{2pt/2pt}c}
			U_{AB}{}^{\bar{A}\bar{B}}(y)		&	0		\\\hdashline[2pt/2pt]
			0							&	(U^{-1})^{AB}{}_{\bar{A}\bar{B}}(y)	\\
		\end{array}
		\right)\,,
\end{align}
with 
\begin{equation}	\label{eq: twist28}
	U_{AB}{}^{\bar{A}\bar{B}}(y)=2\,U_{[A}{}^{\bar{A}}(y)U_{B]}{}^{\bar{B}}(y)\,.
\end{equation}

The $S^7$ reduction \cite{deWit:1986oxb} can be described via \eqref{eq: SSAnsatz} \cite{Hohm:2014qga} in terms of the scaling function
\begin{equation}	\label{eq: rhoS7}
	\rho=(1-g^2\vert y\vert^2)^{1/4}	\,.
\end{equation}
The components of the SL$(8,\mathbb{R})$ matrix $U_{A}{}^{\bar{A}}(y)$ in \eqref{eq: twist28} under the splitting \eqref{eq: E77fundintoGL7} for local and global indices are
\begin{equation}	\label{eq: SL8twistS7}
	\begin{aligned}
		U_i{}^{\bar j}&=(1-g^2\vert y\vert^2)^{-1/8}\big(\delta_i^j+g^2y^i\, y^j\, K_1(y)\big)	\,,	\\[5pt]
		U_i{}^{\bar 8}&=g\,(1-g^2\vert y\vert^2)^{3/8}\, y^i\,,		\\[5pt]
		U_8{}^{\bar j}&=g\,(1-g^2\vert y\vert^2)^{3/8}\, y^j K_1(y)	\,,	\\[5pt]
		U_8{}^{\bar 8}&=(1-g^2\vert y\vert^2)^{7/8}	\,,
	\end{aligned}
\end{equation}
with
\begin{equation}	\label{eq: hyperS7}
	K_1(y)=-{}_2F_1\big(1,3,\tfrac12,1-g^2\vert y\vert^2\big)\,,
\end{equation}
and $\vert y\vert^2=\delta_{ij}y^i\,y^j$. In terms of these coordinates, the round metric obtained through \eqref{eq: GL7dictScalar} by setting the scalars to zero reads
\begin{equation}	\label{eq: roundS7}
	ds^2(S^7_{\rm round})=\delta_{ij}dy^i\,dy^j+\frac{g^{2}(\delta_{ij}y^i\, dy^j)^2}{1-g^{2}\vert y\vert^2} \,.
\end{equation}

The uplift of the dyonic CSO gaugings of \cite{DallAgata:2011aa} can also be described via \eqref{eq: SSAnsatz} with block diagonal twist matrix \eqref{eq: twist} with \eqref{eq: twist28} \cite{Inverso:2016eet}. For the $[\SO(6)\times{\rm SO(2)}]\ltimes\mathbb{R}^{12}$ gauging, the scaling function reads
\begin{equation}	\label{eq: rhoSO}
	\rho=\hat\rho(\hat y)\tilde\rho(\tilde y)\equiv(1- g^2\vert \hat y\vert^2)^{1/4}(1- m^2\tilde y^2)^{1/4}	\,,
\end{equation}
and the components of the SL$(8,\mathbb{R})$ matrix $U_{A}{}^{\bar{A}}(y)$ under the splitting \eqref{eq: GL5embeddings} for local and global indices are
\begin{equation}	\label{eq: SL8twistSO}
	\begin{alignedat}{2}
		U_{\i}{}^{\bar \j}&=\hat\rho^{-1/2}\tilde\rho^{1/2}\big(\delta_{\i}^{\j}+g^2\,\hat y^{\i}\,\hat y^{\j}\, K_2(\hat y)\big)\,,	\qquad\quad
		&&U_{7}{}^{\bar 7}=\hat\rho^{-1/2}\tilde\rho^{1/2}		\,,	\\[5pt]
		U_{\i}{}^{\bar 6}&=g\,\hat\rho^{3/2}\tilde\rho^{1/2}\, \hat y^{\i}\,,		
		&&U_{7}{}^{\bar 8}=-m\,\,\hat\rho^{-1/2}\tilde\rho^{-3/2}\, \tilde y	\,,	\\[5pt]
		U_6{}^{\bar \j}&=g\,\hat\rho^{3/2}\tilde\rho^{1/2}\, \hat y^{\j}\,  K_2(\hat y)	\,,	
		&&U_{8}{}^{\bar 7}=m\,\hat\rho^{-1/2}\tilde\rho^{-3/2}\, \tilde y	\,,	\\[5pt]
		U_6{}^{\bar 6}&=\hat\rho^{7/2}\tilde\rho^{1/2}	\,,		
		&&U_{8}{}^{\bar 8}=\hat\rho^{-1/2}\tilde\rho^{1/2}\,,
	\end{alignedat}
\end{equation}
with $\vert \hat y\vert^2=\delta_{\i \j}\,\hat y^\i\,\hat y^\j$ and
\begin{equation}	\label{eq: hyperSO}
	K_2(\hat y)=-{}_2F_1\big(1,2,\tfrac12,1-g^2\vert \hat y\vert^2\big)\,.
\end{equation}

The generalised frames \eqref{eq: rhoS7}-\eqref{eq: hyperS7} and \eqref{eq: rhoSO}-\eqref{eq: hyperSO} can be checked to satisfy \eqref{eq: parallelisation} with the embedding tensors corresponding to \eqref{eq: gaugings} via \eqref{eq: embtensor912}.


\section{Supergravity Embeddings}


\subsection{SO(8) gauging on $S^7$}	\label{sec: SO8inS7}

In the following, it is convenient to introduce coordinates $(\mu^{\rm a},\ \phi^{\rm a})$, with $\rm a=1,2,3,4$, as\footnote{These coordinates are related to the ones in \cite{Azizi:2016noi} through the redefinitions in footnote~\ref{ftn: redefinitions}.}
\begin{equation}	\label{eq: coordsS7}
	\begin{alignedat}{4}
		&y^1=g^{-1}\mu_1\cos\,\phi_1\,,	\qquad	&&y^3=g^{-1}\mu_2\cos\,\phi_2\,,	\qquad	&&y^5=g^{-1}\mu_3\cos\,\phi_3\,,	\qquad	&&y^7=g^{-1}\mu_4\cos\,\phi_4\,,	\\
		&y^2=g^{-1}\mu_1\sin\,\phi_1\,,				&&y^4=g^{-1}\mu_2\sin\,\phi_2\,,			&&y^6=g^{-1}\mu_3\sin\,\phi_3\,,
	\end{alignedat}
\end{equation}
constrained by
\begin{equation}	\label{eq: constraintS7}
	\begin{aligned}
		\sum_{\rm a=1}^{4}\mu_{\rm a}^2=1\,.
	\end{aligned}
\end{equation}
In terms of these coordinates, the metric with \eqref{eq: SSAnsatz} and \eqref{eq: SL8twistS7} becomes \cite{Azizi:2016noi}
\begin{equation}	\label{eq: defS7}
	\begin{aligned}
		ds_{11}^2=&\,
		\Xi_1^{1/3}ds_4^{2}
		+g^{-2}\Xi_1^{-2/3}\Big[\sum_{\rm a}Z_a(d\mu_{\rm a}^2+\mu_{\rm a}^2D\phi_{\rm a}^2)				\\[4pt]
		&-2b_2b_3(\mu_{3}^2\mu_{4}^2D\phi_{3}D\phi_{4}+\mu_{1}^2\mu_{2}^2D\phi_{1}D\phi_{2})		\\[4pt]
		&-2 b_1 b_3(\mu_{2}^2 \mu_{3}^2 D\phi_2 D\phi_3+\mu_{1}^2 \mu_{4}^2 D\phi_1 D\phi_4)		\\[4pt]
		&-2 b_1 b_2 (\mu_{1}^2 \mu_{3}^2 D\phi_1 D\phi_3+\mu_{2}^2 \mu_{4}^2 D\phi_2 D\phi_4)		\\[4pt]
		&+
		b_1^2 (\mu_{1} d\mu_{1}+\mu_{2} d\mu_{2})^2
		+
		b_2^2 (\mu_{2} d\mu_{2}+\mu_{3} d\mu_{3})^2
		+
		b_3^2 (\mu_{1} d\mu_{1}+\mu_{3} d\mu_{3})^2
		\Big]\,,
	\end{aligned}
\end{equation}
where
\begin{equation}	\label{eq: XiS7}
	\begin{aligned}
		\Xi_1=&\,
		 \tilde{Y}^2_1 \tilde{Y}^2_2Y^2_3 \,\mu_1^4	+ \tilde{Y}^2_1 Y^2_2 \tilde{Y}^2_3\,\mu_2^4+Y^2_1 Y^2_2 Y^2_3\,\mu_3^4+ Y^2_1 \tilde{Y}^2_2 \tilde{Y}^2_3\,\mu_4^4	\\[4pt]
		&+(Y^2_2 \tilde{Y}^2_2+Y^2_3 \tilde{Y}^2_3) \big(Y^2_1\mu_3^2 \mu_4^2+ \tilde{Y}^2_1\mu_1^2\mu_2^2 \big)	\\[4pt]
		&+(Y^2_1 \tilde{Y}^2_1+Y^2_3 \tilde{Y}^2_3) \big(Y^2_2\mu_2^2\mu_3^2 + \tilde{Y}^2_2\mu_1^2\mu_4^2 \big)	\\[4pt]
		&+(Y^2_1 \tilde{Y}^2_1+Y^2_2 \tilde{Y}^2_2) \big(Y^2_3\mu_1^2\mu_3^2 + \tilde{Y}^2_3 \mu_2^2\mu_4^2 \big)\,,
	\end{aligned}
\end{equation}
and
\begin{equation}	\label{eq: SsS7}
	Z_{\rm a}=\mu^2_{\rm a}+W_{\rm a}\,,
\end{equation}
with
\begin{equation}	\label{eq: WsS7}
	\begin{aligned}
		W_1=Y^2_2 \tilde{Y}^2_3\mu_2^2+Y^2_1 Y^2_2\mu_3^2 + Y^2_1 \tilde{Y}^2_3\mu_4^2\,,
		\qquad
		W_2= \tilde{Y}^2_2 Y^2_3\mu_1^2 + Y^2_1 Y^2_3\mu_3^2 + Y^2_1 \tilde{Y}^2_2\mu_4^2 	\,,	\\[4pt] 
		W_3= \tilde{Y}^2_1 \tilde{Y}^2_2\mu_1^2 + \tilde{Y}^2_1 \tilde{Y}^2_3\mu_2^2 + \tilde{Y}^2_2 \tilde{Y}^2_3\mu_4^2	\,,	
		\qquad
		W_4= \tilde{Y}^2_1 Y^2_3\mu_1^2 + \tilde{Y}^2_1 Y^2_2\mu_2^2 + Y^2_2 Y^2_3\mu_3^2	\,.
	\end{aligned}
\end{equation}
For this gauging, the covariant derivatives on the angles denote the fibering with the four vectors
\begin{equation}	\label{eq: DphiS7}
	D\phi_{\rm a}=d\phi_{\rm a}-g A^{\rm a}\,.
\end{equation}
with $A^{\rm a}$ given in terms of SL$(8,\mathbb{R})$ objects in \eqref{eq: TensorHierarchyA}. The three-form potential can be written in terms of the 4$d$ potentials in \eqref{eq: THcontent} suitably coupled to the $S^7$ coordinates, and an internal contribution as dictated by \eqref{eq: GL7dictScalar}. The link between the tensor hierarchy fields and the sphere coordinates is in fact fixed by their respective ${\rm SL}(8,\mathbb{R})$ structure, and gauge invariance demands that the different terms combine into the field strengths in \eqref{eq: THFieldStrengths} when acted upon by the exterior derivative. Notably, only a subset of the forms in \eqref{eq: THcontent}, dubbed ``restricted tensor hierarchy'' in \cite{Guarino:2015qaa,Varela:2015ywx}, enters the KK Ansatz. For the STU truncation, the eleven-dimensional three-form can be decomposed as simply
\begin{equation}	\label{eq: 3formTotal}
	\hat{A}_\3=\sum_{\rm a=1}^4\Big[-C_{\rm a}\, \mu_{\rm a}^2+\tfrac{1}{2g}\big[B_{\rm a}+\tfrac12A^{\rm a}\wedge\tilde{A}_{\rm a}\big]\wedge d(\mu^2_{\rm a})-\tfrac1{2g^{2}}\tilde{A}_{\rm a}\wedge d(\mu_{\rm a}^2)\wedge D\phi^{\rm a}\Big]+C_\3\,,
\end{equation}
with the overall scaling fixed by the equations of motion. This result agrees with the truncation of \cite{Varela:2015ywx} through \eqref{eq: TensorHierarchyA}--\eqref{eq: TensorHierarchyC420}. The expression for the internal three-form can also be obtained from \eqref{eq: GL7dictScalar} with \eqref{eq: SSAnsatz} and \eqref{eq: SL8twistS7}, and reads
\begin{equation}	\label{eq: 3formInternal}
	C_\3=\frac12\sum_{\rm abc}C_{\rm a,bc}d\mu^{\rm a}\wedge D\phi^{\rm b}\wedge D\phi^{\rm c}\,,
\end{equation}
with components $C_{\rm a,bc}=C_{\rm a,[bc]}$ given by
\begin{equation}	\label{eq: 3formInternalComponents}
	\begin{aligned}
		C_{\rm a, 12}\,d\mu^{\rm a}&=\frac{-b_1}{2\,g^3\,\Xi_1}\Big[\mu_1^2\,W_2\, d(\mu_2^2)-\mu_2^2\,W_1\, d(\mu_1^2)
			+\mu_1^2\mu_2^2\big(Y_2^2\tilde{Y}_2^2\, d\alpha_{23}-Y_3^2\tilde{Y}_3^2\, d\alpha_{13}\big)	\Big]\,,	\\[4pt]
		C_{\rm a, 34}\,d\mu^{\rm a}&=\frac{-b_1}{2\,g^3\,\Xi_1}\Big[\mu_3^2\,W_4\, d(\mu_4^2)-\mu_4^2\,W_3\, d(\mu_3^2)
			-\mu_3^2\mu_4^2\big(Y_2^2\tilde{Y}_2^2\, d\alpha_{23}+Y_3^2\tilde{Y}_3^2\, d\alpha_{13}\big)	\Big]\,,	\\[4pt]
		C_{\rm a, 23}\,d\mu^{\rm a}&=\frac{-b_2}{2\,g^3\,\Xi_1}\Big[\mu_2^2\,W_3\, d(\mu_3^2)-\mu_3^2\,W_2\, d(\mu_2^2)
			+\mu_2^2\mu_3^2\big(Y_3^2\tilde{Y}_3^2\, d\alpha_{13}-Y_1^2\tilde{Y}_1^2\, d\alpha_{12}\big)	\Big]\,,	\\[4pt]
		C_{\rm a, 14}\,d\mu^{\rm a}&=\frac{-b_2}{2\,g^3\,\Xi_1}\Big[\mu_1^2\,W_4\, d(\mu_4^2)-\mu_4^2\,W_1\, d(\mu_1^2)
			-\mu_1^2\mu_4^2\big(Y_1^2\tilde{Y}_1^2\, d\alpha_{12}+Y_3^2\tilde{Y}_3^2\, d\alpha_{13}\big)	\Big]\,,	\\[4pt]
		C_{\rm a, 13}\,d\mu^{\rm a}&=\frac{-b_3}{2\,g^3\,\Xi_1}\Big[\mu_1^2\,W_3\, d(\mu_3^2)-\mu_3^2\,W_1\, d(\mu_1^2)
			+\mu_1^2\mu_3^2\big(Y_2^2\tilde{Y}_2^2\, d\alpha_{23}-Y_1^2\tilde{Y}_1^2\, d\alpha_{12}\big)	\Big]\,,	\\[4pt]
		C_{\rm a, 24}\,d\mu^{\rm a}&=\frac{-b_3}{2\,g^3\,\Xi_1}\Big[\mu_2^2\,W_4\, d(\mu_4^2)-\mu_4^2\,W_2\, d(\mu_2^2)
			-\mu_2^2\mu_4^2\big(Y_1^2\tilde{Y}_1^2\, d\alpha_{12}+Y_2^2\tilde{Y}_2^2\, d\alpha_{23}\big)	\Big]\,,	\\[4pt]
	\end{aligned}
\end{equation}
and shorthands
\begin{equation}	\label{eq: alphas}
	\alpha_{12}=\mu_1^2+\mu_2^2\,,	\qquad
	\alpha_{23}=\mu_2^2+\mu_3^2\,,	\qquad
	\alpha_{34}=\mu_3^2+\mu_4^2\,,	\qquad
	\rm{etc.}
\end{equation}
This result matches (4.19) of \cite{Azizi:2016noi} upon making the identifications in footnote~\ref{ftn: redefinitions}.
The eleven-dimensional four-form is then produced by the exterior derivative of \eqref{eq: 3formTotal}, which by using \eqref{eq: THFieldStrengths} can be given as
\begin{equation}	\label{eq: 4formTotal}
	\begin{aligned}
		\hat{F}_\4=&\sum_{\rm a=1}^4\Big[-H_{\4\rm a}\, \mu_{\rm a}^2+\tfrac1{2g}H_{\3\rm a}\wedge d(\mu^2_{\rm a})+\tfrac1{2g^2}\tilde{H}_{\2\rm a}\wedge d(\mu_{\rm a}^2)\wedge D\phi^{\rm a}\Big]\\[5pt]
		&+g\sum_{\rm abc}C_{\rm a,bc}d\mu^{\rm a}\wedge D\phi^{\rm b}\wedge H_\2^{\rm c}
		+\frac12\sum_{\rm abcd}\partial_dC_{\rm a,bc}d\mu^{\rm d}\wedge d\mu^{\rm a}\wedge D\phi^{\rm b}\wedge D\phi^{\rm c}	\\
		&\qquad+\frac12\sum_{\rm abc}\partial_\mu C_{\rm a,bc}dx^\mu\wedge d\mu^{\rm a}\wedge D\phi^{\rm b}\wedge D\phi^{\rm c}\,,
	\end{aligned}
\end{equation}
and thus making use of only the original fields appearing in the $\mathcal{N}=8$ action and their derivatives through the duality relations \eqref{eq: duality2form}--\eqref{eq: duality4form}, with the last equation particularised for this gauging to
\begin{equation}	\label{eq: duality4formSO8}
	\begin{aligned}
		H_{\41}&=2g(\tilde{Y}_1^2+\tilde{Y}_2^2+Y_3^2)\vol_4\,,	\qquad\quad
		H_{\42}=2g(\tilde{Y}_1^2+Y_2^2+\tilde{Y}_3^2)\vol_4\,,	\\[5pt]
		H_{\43}&=2g(Y_1^2+Y_2^2+Y_3^2)\vol_4\,,	\qquad\quad
		H_{\44}=2g(Y_1^2+\tilde{Y}_2^2+\tilde{Y}_3^2)\vol_4\,.
	\end{aligned}
\end{equation}

\subsubsection*{Singular limit}

Scaling the fields and couplings as in \eqref{eq: scalings} and \eqref{eq: scalingvecs}, the configuration remains finite up to a trombone scaling if we also transform the internal coordinates as
\begin{equation}	\label{eq: muscaling}
	\mu_4\mapsto e^{-k}\mu_4	\,,
\end{equation}
with the coordinates $(\mu_{\rm\bar{a}},\ \phi_{\rm\bar{a}},\ \phi_4)$ invariant. Doing so, the warping \eqref{eq: XiS7} factorises into
\begin{equation}
	\Xi_1\mapsto\bar\Xi_1=e^k\, H\, \Xi_2	\,,
\end{equation}
with
\begin{equation}	\label{eq: XiS5xS1}
	H=\tilde{Y}_2^2\mu_1^2+\tilde{Y}_3^2\mu_2^2+Y_1^2\mu_3^2\,,
	\qquad\qquad
	\Xi_2=W_4 \,.
\end{equation}
Accordingly, the metric becomes
\begin{align}	\label{eq: scaledMetric}
	ds_{11}^2\mapsto d\bar{s}_{11}^2=&\,
	e^{k/3}\Big\{(H\,\Xi_2)^{1/3}ds_4^{2}+g^{-2}\, (H\,\Xi_2)^{-2/3}\Big[\,\Xi_2(d\mu_{4}^2+\mu_{4}^2d\phi_{4}^2)	\nonumber\\
	&\qquad+H\big(Y_2^2(d\mu_{1}^2+\mu_{1}^2D\phi_{1}^2)+Y_3^2(d\mu_{2}^2+\mu_{2}^2D\phi_{2}^2)+\tilde{Y}_1^2(d\mu_{3}^2+\mu_{3}^2D\phi_{3}^2)\big)	\nonumber\\
	&\qquad\quad-\big(b_2\mu_1^2D\phi_1+b_3\mu_2^2D\phi_2+b_1\mu_3^2D\phi_3\big)^2	
	\Big]\Big\}\,,
\end{align}
and the three-form components \eqref{eq: 3formInternalComponents} transform as
\begin{equation}
	C_{a,bc}d\mu^a \mapsto e^{k/2}\bar{C}_{a,bc}d\mu^a\,,
\end{equation}
with
\begin{equation}	\label{eq: 3formLimit}
	\begin{aligned}
		\bar C_{\rm a, 12}\,d\mu^{\rm a}=\frac{b_1\,\mu_1\mu_2}{g^3\,\Xi_2}\big(Y_2^2\mu_2d\mu_1&-Y_3^2\mu_1d\mu_2\big)\,,	\qquad
		\bar C_{\rm a, 23}\,d\mu^{\rm a}=\frac{b_2\,\mu_2\mu_3}{g^3\,\Xi_2}\big(Y_3^2\mu_3d\mu_2-\tilde{Y}_1^2\mu_2d\mu_3\big)\,,	\\[5pt]
		&\bar C_{\rm a, 13}\,d\mu^{\rm a}=\frac{b_3\,\mu_1\mu_3}{g^3\,\Xi_2}\big(Y_2^2\mu_3d\mu_1-\tilde{Y}_1^2\mu_1d\mu_3\big)\,,
	\end{aligned}
\end{equation}
and
\begin{equation}	\label{eq: 3formLimitII}
	\begin{aligned}
		\bar C_{\rm a, 41}\,d\mu^{\rm a}=\frac{b_2}{2g^3H}\mu_1^2\, d(\mu_4^2)\,,&	\qquad\qquad
		\bar C_{\rm a, 42}\,d\mu^{\rm a}=\frac{b_3}{2g^3H}\mu_2^2\, d(\mu_4^2)\,,	\\[5pt]
		\bar C_{\rm a, 43}\,d\mu^{\rm a}&=\frac{b_1}{2g^3H}\mu_3^2\, d(\mu_4^2)\,.
	\end{aligned}
\end{equation}
Therefore, the four-form \eqref{eq: 4formTotal} becomes
\begin{align}	\label{eq: scaled4formTotal}
		\hat{F}_\4\mapsto\bar{\hat{F}}_\4&=e^{k/2}\Big\{\sum_{\rm \bar a=1}^3\Big[-\bar H_{\4\rm \bar a}\, \mu_{\rm \bar a}^2+\tfrac1{2g}H_{\3\rm \bar a}\wedge d(\mu^2_{\rm \bar a})+\tfrac1{2g^2}\tilde{H}_{\2\rm \bar a}\wedge d(\mu_{\rm \bar a}^2)\wedge D\phi^{\rm \bar a}\Big]	\nonumber\\[5pt]
		&\qquad\qquad\qquad
		+\frac1{2}\sum_{\rm \bar a\bar b\bar c}d\Big[\bar C_{\rm \bar a,\bar b\bar c}d\mu^{\rm \bar a}\wedge D\phi^{\rm \bar b}\wedge D\phi^{\rm \bar c}\Big]	\\[5pt]
		&\qquad\qquad+\tfrac1{2g^2}\tilde{H}_{\24}\wedge d(\mu_{4}^2)\wedge d\phi^{4}-\sum_{\rm \bar a}d\Big[\bar C_{\rm 4,\bar a4} D\phi^{\rm \bar a}\Big]\wedge d\mu^4\wedge d\phi^4\Big\}	\,,	\nonumber
\end{align}
with the $\bar H_{\4\rm \bar a}$ in \eqref{eq: H4scalings} reading
\begin{equation}	\label{eq: hshort}
	\bar H_{1}=2g(\tilde{Y}_1^2+Y_3^2)\vol_4\,,	\qquad
	\bar H_{2}=2g(\tilde{Y}_1^2+Y_2^2)\vol_4\,,	\qquad
	\bar H_{3}=2g(Y_2^2+Y_3^2)\vol_4\,.
\end{equation}
%

\subsubsection*{IIA reduction and dualisation to IIB}
The metric and four-form in \eqref{eq: scaledMetric} and \eqref{eq: scaled4formTotal} formally describe a warped compactification on $S^5\times\mathbb{R}^2$, with the $\mathbb{R}^2$ factor parameterised by $(\mu_4,\, \phi_4)$ and the sphere by $(\mu_{\rm \bar{a}},\, \phi_{\rm \bar{a}})$ satisfying
\begin{equation}	\label{eq: constraintS5xS1}
	\begin{aligned}
		\sum_{\rm \bar{a}=1}^3\mu_{\rm \bar{a}}^2=1\,,
	\end{aligned}
\end{equation}
which follows from \eqref{eq: constraintS7} after taking the $k\to\infty$ limit on \eqref{eq: muscaling}. Introducing coordinates $z_{1,2}$ as
\begin{equation}
	z_1=g^{-1}\mu_4\cos\phi_4\,,	\qquad\quad
	z_2=g^{-1}\mu_4\sin\phi_4\,,
\end{equation}
we can promote the $\mathbb{R}^2$ factor into a two-torus by imposing $z_{1,2}\sim z_{1,2}+2\pi R_{1,2}$. The eleven-dimensional configuration, \eqref{eq: scaledMetric} and \eqref{eq: scaled4formTotal}, can then be interpreted as a type IIA supergravity solution upon reducing on one of the circles, say $z_2$. The resulting IIA geometry has an $S^1$ factor that allows one to perform a T-duality transformation and in this way obtain a solution of type IIB supergravity.

To ease our notation, we introduce the following shorthands
\begin{align}
	X_\4	&=\sum_{\rm \bar a=1}^3\Big[-\bar H_{\4\rm \bar a}\, \mu_{\rm \bar a}^2+\tfrac1{2g}H_{\3\rm \bar a}\wedge d(\mu^2_{\rm \bar a})+\tfrac1{2g^2}\tilde{H}_{\2\rm \bar a}\wedge d(\mu_{\rm \bar a}^2)\wedge D\phi^{\rm \bar a}\Big] \nonumber\\[3pt]	
		&\qquad\quad+\frac1{2}\sum_{\rm \bar a\bar b\bar c}d\Big[\bar C_{\rm \bar a,\bar b\bar c}d\mu^{\rm \bar a}\wedge D\phi^{\rm \bar b}\wedge D\phi^{\rm \bar c}\Big]\,,	\nonumber\\[7pt]
	X_\2 &=\tilde{H}_{\24}+d\Big[\frac{1}{gH}\big(b_2\,\mu_1^2 D\phi_1+b_3\,\mu_2^2 D\phi_2 +b_1\,\mu_3^2 D\phi_3\big)\Big] \label{eq: X2}\,,
\end{align}
such that the eleven-dimensional four-form can be written as
\begin{equation}	\label{eq: fourformconv}
	\bar {\hat {F}}_\4= X_\4+ X_\2 \wedge dz_1 \wedge dz_2\,.
\end{equation}
Both $X_\2$ and $X_\4$ are exact, and a representative potential for $X_\2$ can be read off from \eqref{eq: X2} to be
\begin{equation}	\label{eq: IIBKKvector}
	\mathcal{A} = \tilde{A}_4+\frac{1}{gH}\Big[b_2\,\mu_1^2 D\phi_1+b_3\,\mu_2^2 D\phi_2+b_1\,\mu_3^2 D\phi_3\Big]\,.
\end{equation}

Following the conventions stated in appendix~\ref{app: conventions}, the type IIA configuration resulting from reduction of \eqref{eq: scaledMetric} and \eqref{eq: fourformconv} on $z_2$ is
\begin{align}	\label{eq: redIIA}
	e^{\phi_{\rm IIA}} &= \frac{\Xi_2^{1/4}}{\sqrt{H}}\,,\quad  
	H_{\3 \rm IIA} = X_\2 \wedge dz_1\,,\quad 
	F_{\2 \rm IIA} = 0\,,\quad 
	F_{\4 \rm IIA} = X_\4\,, \nonumber\\[5pt]
	ds_{\rm IIA}^2 &= H^{1/4}\Xi_2^{3/8} ds^2_4 +g^{-2}\,H^{-3/4}\Xi_2^{-5/8}\Big[g^2\,\Xi_2\,dz_1^2-(b_2\mu_1^2  D\phi_1 +b_3\mu_2^2D\phi_2+b_1\mu_3^2 D\phi_3)^2		\nonumber\\[5pt]
	& \quad\quad +H\big(Y_2^2(d\mu_1^2+\mu_1^2D\phi_1^2)+Y_3^2(d\mu_2^2+\mu_2^2D\phi_2^2)+\tilde{Y}_1^2(d\mu_3^2+\mu_3^2D\phi_3^2)\big)\Big]\,,
\end{align}
and the type IIB solution obtained by T-dualising along the $z_1$ direction employing the relations \eqref{eq: TdualScalars}-\eqref{eq: TdualForms} reads
\begin{align}	\label{eq: IIBfrom11D}
	&\phi_{\rm IIB}=\chi=0\,,	\nonumber\\[7pt]	
	&ds_{\rm IIB}^2 = \Xi_2^{1/2} ds^2_4 +g^{-2}\,\Xi_2^{-\frac12}\big[Y_2^2(d\mu_1^2+\mu_1^2D\phi_1^2)+Y_3^2(d\mu_2^2+\mu_2^2D\phi_2^2)+\tilde{Y}_1^2(d\mu_3^2+\mu_3^2D\phi_3^2)
	\nonumber\\[5pt]	
	& \quad \quad\quad -H^{-1} (b_2\mu_1^2  D\phi_1 +b_3\mu_2^2D\phi_2+b_1\mu_3^2 D\phi_3)^2 + g^2H(dz_1+\mathcal{A})^2\big]\,, 	\nonumber\\[7pt]
	&H_{\3 \rm IIB} = F_{\3 \rm IIB} = 0\,,	\nonumber\\[5pt]	
	&F_{\5 \rm IIB} = (1+*_{10})\big[X_\4\wedge \big(dz_1+\mathcal{A}\big)\big]\,.	
\end{align}
%

\subsection{$[\SO(6)\times\SO(2)]\ltimes\mathbb{R}^{12}$ gauging on $S^5\times S^1$}

The type IIB configurations that uplift from $D=4$ gauged supergravity with gauging specified by $(\theta_{\sst{(6c)}}, \xi_{\sst{(6c)}})$ in \eqref{eq: gaugings} can also be obtained from \eqref{eq: SSAnsatz}, by employing the twist in \eqref{eq: rhoSO}--\eqref{eq: hyperSO} and the ExFT dictionary. For the STU sector, this configuration is related to the singular limit of the $S^7$ configurations in M-theory in the last section.
To accomplish this, we use the coordinates $(\mu_{\rm \bar{a}},\, \phi_{\rm \bar{a}})$ above
related to $y^i$ as
\begin{equation}	\label{eq: coordsS5xS1}
	\begin{alignedat}{3}
		&\hat y^1=g^{-1}\mu_1\cos\,\phi_1\,,	\qquad	&&\hat y^3=g^{-1}\mu_2\cos\,\phi_2\,,	\qquad	&&\hat y^5=g^{-1}\mu_3\cos\,\phi_3\,,	\\
		&\hat y^2=g^{-1}\mu_1\sin\,\phi_1\,,			&&\hat y^4=g^{-1}\mu_2\sin\,\phi_2\,,	
	\end{alignedat}
\end{equation}
and constrained by \eqref{eq: constraintS5xS1}. A coordinate $z$ can also be introduced such that
\begin{equation}
	\tilde{y}=m^{-1}\sin mz\,,
\end{equation}
which reduces to $\tilde{y}=z$ in the vanishing $m$ limit. The metric obtained from \eqref{eq: GL6xSL2dictScalar} reads
\begin{equation}	\label{eq: defS5xS1z}
	\begin{aligned}
		d\hat{s}_{10}^2=&\,
		\Xi_2^{1/2}ds_4^{2}
		+g^{-2}\Xi_2^{-1/2}\Big[
		Y^2_2 (d\mu_1^2+\mu_1^2 D\phi_1^2)+Y^2_3 (d\mu_2^2+\mu_2^2 D\phi_2^2)+\tilde{Y}^2_1(d\mu_3^2+\mu_3^2 D\phi_3^2)		\\
		&\qquad\qquad\qquad\qquad\enspace-H^{-1}\big(b_2\mu_1^2D\phi_1+b_3\mu_2^2D\phi_2+b_1\mu_3^2D\phi_3\big)^2		\\
		&\qquad\qquad\qquad\qquad\quad+H\Big(g\,dz+g\tilde{A}_{4}+H^{-1}\big(b_2\mu_1^2D\phi_1+b_3\mu_2^2D\phi_2+b_1\mu_3^2D\phi_3\big)\Big)^2
		\Big]\,,
	\end{aligned}
\end{equation}
where $\Xi_2$ and $H$ are given in \eqref{eq: XiS5xS1} and, for this gauging, the covariant derivatives on the angles denote the fibering with the four vectors
\begin{equation}	\label{eq: DphiS5xS1}
	D\phi_{\rm \bar{a}}=d\phi_{\rm \bar{a}}-gA^{\rm \bar{a}}\,,
\end{equation}
with $A^{\rm \bar{a}}=\{A^{12}\,,\ A^{34}\,,\ A^{56}\}$ in terms of SL$(6,\mathbb{R})$ indices. Note that for these flat indices, ${\rm SL}(6,\mathbb{R})\subset{\rm SL}(8,\mathbb{R})$.
In this sector, the axiodilaton vanishes
\begin{equation}	\label{eq: axiodilatonSO2}
	\Phi=0\,,	\qquad\quad
	C_{0}=0\,,
\end{equation}
as well as the two-form potentials

\begin{equation}	\label{eq: 2formsSO2}
	C_{\2}{}^a=0	\,.
\end{equation}
Finally, the four-form gauge potential can also be derived from the ExFT dictionary. The purely internal contribution can be obtained from \eqref{eq: GL6xSL2dictScalar} to be
\begin{equation}	\label{eq: 4formSO2}
	\begin{aligned}
		C_{\4}&=-g^{-4}\mu_1\mu_2(1+K_2)\frac{\cot\phi_3}{\mu_3^2}d\mu_1\wedge d\mu_2\wedge d\phi_1\wedge d\phi_2	\\[5pt]
		&\quad-g^{-4}\mu_1\mu_2\Big[K_2(\mu_2d\mu_1-\mu_1d\mu_2)+\Xi_2^{-1}\tilde{Y}_1^2(Y_2^2\mu_2d\mu_1-Y_3^2\mu_1d\mu_2)\Big]\wedge d\phi_1\wedge d\phi_2\wedge d\phi_3	\\[5pt]
		&\qquad+\tfrac1{2g^{3}}\Xi_2^{-1}\Big[b_1\big(Y_2^2\mu_2^2d(\mu_1^2)-Y_3^2\mu_1^2d(\mu_2^2)\big)\wedge d\phi_1\wedge d\phi_2	\\[5pt]
		&\qquad\qquad\qquad+b_2\big(Y_2^2\mu_3^2d(\mu_1^2)-\tilde{Y}_1^2\mu_1^2d(\mu_3^2)\big)\wedge d\phi_1\wedge d\phi_3			\\[5pt]
		&\qquad\qquad\qquad\quad+b_3\big(Y_3^2\mu_3^2d(\mu_2^2)-\tilde{Y}_1^2\mu_2^2d(\mu_3^2)\big)\wedge d\phi_2\wedge d\phi_3\Big]\wedge dz	\,.
	\end{aligned}
\end{equation}
We recognise that, stripped of $dz$, the contributions in the last three lines in \eqref{eq: 4formSO2} precisely match the coefficients $\bar{C}_{\rm \bar a,\bar b\bar c}$ in \eqref{eq: 3formLimit}, and the derivatives of the first two lines are dual to the contributions in \eqref{eq: 3formLimit} given by the forms in the tensor hierarchy. Therefore, the associated self-dual five-form field strength is given by
\begin{equation}	\label{eq: fiveformSO6SO2}
	\begin{aligned}
		\hat{F}_{\5} &= (1+*_{10})\Big\{\Big[\sum_{\rm \bar a=1}^3\Big(-H_{\4\rm \bar a}\, \mu_{\rm \bar a}^2+\tfrac1{2g}H_{\3\rm \bar a}\wedge d(\mu^2_{\rm \bar a})+\tfrac1{2g^2}\tilde{H}_{\2\rm \bar a}\wedge d(\mu_{\rm \bar a}^2)\wedge D\phi^{\rm \bar a}\Big)			\\[5pt]
		&\qquad\qquad\qquad+\frac1{2}\sum_{\rm \bar a\bar b\bar c}d\Big(\bar C_{\rm \bar a,\bar b\bar c}d\mu^{\rm \bar a}\wedge D\phi^{\rm \bar b}\wedge D\phi^{\rm \bar c}\Big)\Big]	\\[5pt]
		&\qquad\qquad\qquad\qquad 
		\wedge \Big[dz+\tilde{A}_{4}+\frac1{gH}\Big(b_2\,\mu_1^2 D\phi_1+b_3\,\mu_2^2 D\phi_2 +b_1\,\mu_3^2 D\phi_3\Big)\Big]\Big\}\,,	
	\end{aligned}
\end{equation}
with $\bar C_{\rm \bar a,\bar b\bar c}$ given in \eqref{eq: 3formLimit}, and the field strengths in the tensor hierarchy given by \eqref{eq: duality2form}, \eqref{eq: duality3form}  and \eqref{eq: duality4form} with $x=0$ and $\tilde{x}=1$, which coincide with the ones in \eqref{eq: hshort}. The uplift formulae \eqref{eq: defS5xS1z}--\eqref{eq: fiveformSO6SO2} precisely match the type IIB configuration obtained in \eqref{eq: IIBfrom11D} if one identifies the angles on $S^1$.

Even though the gauge coupling $m$ does not enter into the Kaluza-Klein Ans\"atze for the type IIB bosonic fields, it mediates the relation between ten- and four-dimensional spinors, as expected from the different fermion couplings observed in four dimensions.

\vspace{8pt}
To perform a thorough check of the Ansatz in \eqref{eq: defS5xS1z}--\eqref{eq: fiveformSO6SO2}, we will restrict our attention to a simpler truncation that identifies the fields as 
\begin{equation}	\label{eq: 3+1truncation}
	\varphi_1=-\varphi_2=-\varphi_3=\tfrac1{\sqrt{3}}\varphi\,,	\qquad\quad
	\chi_{1,2,3}=\tfrac1{\sqrt{3}}\chi\,,								\qquad\quad
	A^{1,2,3}_\1=\tfrac1{\sqrt3}A_\1\,,	
\end{equation}
and leaves $A^{4}_\1$ unfixed. 
For the $\SO(6)\ltimes\mathbb{R}^{12}$ gauging, this theory can be obtained as a circle reduction of the $\mathcal{N}=4$ ${\rm SU(2)\times U(1)}$ gauged theory in $D=5$ \cite{Romans:1985ps,Lu:1999bw} truncated so that the five-dimensional scalar vanishes and the gauge group is reduced as ${\rm U(1)}\subset {\rm U(1)\times U(1)}\subset{\rm SU(2)\times U(1)}$. After the circle reduction, one can identify $A^4_\1$ with the dual of the KK vector.
In the following, we only consider configurations with $\chi=0$, which per \eqref{eq: THBianchi} cannot be dyonically charged so as to guarantee $H_\2^{\rm a}\wedge H_\2^{\rm b}=0$.
The Lagrangian \eqref{eq: STULagrangian} for the $\SO(6)\ltimes\mathbb{R}^{12}$ and $[\SO(6)\times\SO(2)]\ltimes\mathbb{R}^{12}$ gaugings then becomes
\begin{equation}	\label{eq: 111Lagrangian}
	e^{-1}\mathcal{L}=R-\tfrac12(\partial\varphi)^2-\tfrac14e^{a\varphi}H^2-\tfrac14e^{-\varphi/a}(H^4)^2+12g^2e^{-a\varphi}\,.
\end{equation}
with $H_{\2}=dA_\1$. To make contact with the STU models in \eqref{eq: STULagrangian}, the dilaton coupling needs to be set as $a=1/\sqrt3$, but we find it convenient to keep it unspecified at the four-dimensional level.

For $a=1/\sqrt3$, we can embed any solution of the $4d$ theory in ten dimensions. From \eqref{eq: defS5xS1z}, the metric reads
\begin{equation}	\label{eq: metricIIB3+1}
	\begin{aligned}
		d\hat{s}_{10}^2=&\,
		e^{\frac{-\varphi}{\sqrt3}}ds_4^{2}
		+g^{-2}\Big[
		d\mu_1^2+d\mu_2^2+d\mu_3^2+\mu_1^2 D\phi_1^2+\mu_2^2 D\phi_2^2+\mu_3^2 D\phi_3^2\Big]+e^{\frac{2\varphi}{\sqrt3}}\big(dz+\tilde{A}_{4}\big)^2\,,
	\end{aligned}
\end{equation}
and the tensor hierarchy fields in \eqref{eq: TH3+1} allow us to write the five-form as
\begin{equation}	\label{eq: fiveformIIB3+1}
	\begin{aligned}
		\hat{F}_{\5} &=(1+*_{10})\Big\{\Big[-4g\, e^{-\frac{\varphi}{\sqrt3}}\,\vol_4
			-\tfrac1{2\sqrt3g^2}e^{\frac{\varphi}{\sqrt3}}\,*H_\2\wedge \sum_{\rm \bar a=1}^3\big(d(\mu_{\rm \bar a}^2)\wedge D\phi^{\rm \bar a}\big)\Big]\wedge \big(dz+\tilde{A}_{4}\big)	\Big\}
			\\[5pt]
			&=\Big(-4g\, e^{-\frac{\varphi}{\sqrt3}}\,\vol_4
			-\tfrac1{2\sqrt3g^2}e^{\frac{\varphi}{\sqrt3}}\,*H_\2\wedge \sum_{\rm \bar a=1}^3\big(d(\mu_{\rm \bar a}^2)\wedge D\phi^{\rm \bar a}\big)\Big)\wedge \big(dz+\tilde{A}_{4}\big) \\
			&\qquad-
			4g^{-4}\,\mu_1\mu_2d\mu_{1}\wedge d\mu_2\wedge D\phi^{1}\wedge D\phi^{2}\wedge D\phi^{3}	\\
			&\qquad\quad+\tfrac{1}{\sqrt3g^3}\,H_\2\wedge \Big[\mu_1\mu_2(\mu_1d\mu_2-\mu_2d\mu_1)\wedge D\phi^1\wedge D\phi^2\\
			&\qquad\qquad+\mu_2\mu_3(\mu_2d\mu_3-\mu_3d\mu_2)\wedge D\phi^2\wedge D\phi^3+\mu_1\mu_3(\mu_1d\mu_3-\mu_3d\mu_1)\wedge D\phi^1\wedge D\phi^3\Big]\,,
	\end{aligned}
\end{equation}
using the duality relations \eqref{eq: duality2form}--\eqref{eq: duality4form}, that reduce to
\begin{equation}	\label{eq: TH3+1}
	\begin{aligned}
		\tilde{H}_{\2\,1,2,3}&=-\tfrac1{\sqrt3}\,e^{\frac{\varphi}{\sqrt3}}\,*H_\2\,,		\qquad\quad\; 
		\tilde{H}_{\2\,4}=-e^{-\sqrt3\varphi}\,*H_{\2}^4\,,							\\[5pt]
		H_{\3\,1,2,3}&=-\tfrac{1}{2\sqrt3}*d\varphi\,,							\qquad\qquad\quad\,\;  
		H_{\3\,4}=\tfrac{\sqrt3}{2}*d\varphi\,,									\\[5pt]
		H_{\4\,1,2,3}&=4g\, e^{-\frac{\varphi}{\sqrt3}}\,\vol_4\,,					\qquad\quad\enspace\quad\, 
		H_{\4\,4}=6g\, e^{\frac{\varphi}{\sqrt3}}\,\vol_4\,,							
	\end{aligned}
\end{equation}
upon using the truncation \eqref{eq: 3+1truncation} with $\chi=0$. 
From \eqref{eq: IIBpaction}, the Bianchi identity for the five-form and Einstein equations are the only equations to be checked in ten dimensions.
Since the axiodilation and two-forms in \eqref{eq: axiodilatonSO2} and \eqref{eq: 2formsSO2} are zero, the equation of motion for the type IIB five-form amounts to demanding that $F_{\5}$ be closed. In \eqref{eq: fiveformIIB3+1}, it is straightforward to see that this is in turn an immediate consequence of the four-dimensional Bianchi identities and equations of motion from \eqref{eq: 111Lagrangian}.

For vanishing axiodilaton and two-forms, Einstein equations in 10$d$ in turn reduce to
\begin{equation}	\label{eq: EinsteinIIB}
	\hat{G}=\hat{T}=-\frac1{480}\Big[\hat{F}_{\hat{\mu}\hat{\rho}_1\hat{\rho}_2\hat{\rho}_3\hat{\rho}_4}\hat{F}_{\hat{\nu}}{}^{\hat{\rho}_1\hat{\rho}_2\hat{\rho}_3\hat{\rho}_4}
	-\frac1{10}\hat{g}_{\hat{\mu}\hat{\nu}}\hat{F}_{\hat{\rho}_1\hat{\rho}_2\hat{\rho}_3\hat{\rho}_4\hat{\rho}_5}\hat{F}^{\hat{\rho}_1\hat{\rho}_2\hat{\rho}_3\hat{\rho}_4\hat{\rho}_5}\Big]v^{\hat{\mu}}v^{\hat{\nu}}\,,
\end{equation}
with $\hat{G}=\hat{G}_{\hat{\mu}\hat{\nu}}v^{\hat{\mu}}v^{\hat{\nu}}$ the Einstein tensor for $\hat{g}_{10}$. We find it convenient to expand our tensors in the one-form basis
\begin{equation}	\label{eq: basis10}
	v^{\hat{\mu}}=\big\{dx^\mu,\ v^m\big\}=\Big\{dx^\mu,\ d\alpha,\ d\beta,\ D\phi^1,\ D\phi^2,\ D\phi^3,\ Dz\Big\}\,,
\end{equation}
where $dx^\mu$ is a coordinate basis for the four-dimensional spacetime and the angles on the sphere are given by
\begin{equation}	
	\mu_1=\sin\alpha\cos\beta\,,	\qquad\quad
	\mu_2=\sin\alpha\sin\beta\,,	\qquad\quad
	\mu_3=\cos\alpha\,.
\end{equation}
In this basis, the Einstein tensor reads
\begin{subequations}
\begin{align}
	\hat{G}_{\mu\nu}&=G_{\mu\nu}-\tfrac16e^{\frac{\varphi}{\sqrt3}}\Big(H_{\mu\rho}H_{\nu}{}^{\rho}-\tfrac14g_{\mu\nu}H_{\rho\sigma}H^{\rho\sigma}\Big)-\tfrac12e^{\sqrt3\varphi}\Big(\tilde H_{4\mu\rho}\tilde H_{4\nu}{}^{\rho}-\tfrac14g_{\mu\nu}\tilde H_{4\rho\sigma}\tilde H_4^{\rho\sigma}\Big)	\nonumber\\[4pt]
				&\qquad-\tfrac12\big(\partial_{\mu}\varphi\partial_{\nu}\varphi-\tfrac12g_{\mu\nu}\partial_{\rho}\varphi\partial^{\rho}\varphi\big)-10g^2e^{-\frac{\varphi}{\sqrt3}}g_{\mu\nu}\,,	\\[8pt]
	\hat{G}_{\mu m}v^m&=\tfrac1{\sqrt3g}\nabla^{\nu}\big(e^{\frac{\varphi}{\sqrt3}}H_{\nu\mu}\big)\big(\mu_1^2D\phi_1+\mu_2^2D\phi_2+\mu_3^2D\phi_3\big)-\nabla^{\nu}\big(e^{\sqrt3\varphi}\tilde H_{4\nu\mu}\big)Dz\,,	\\[8pt]
	\hat{G}_{mn}v^mv^n&=\tfrac12e^{\frac{\varphi}{\sqrt3}}\Big[g^{\mu\nu}G_{\mu\nu}+\tfrac1{12}e^{\frac{\varphi}{\sqrt3}}H_{\mu\nu}H^{\mu\nu}+\tfrac14e^{\sqrt3\varphi}\tilde H_{4\mu\nu}\tilde H_4^{\mu\nu}-\tfrac1{\sqrt3}\square\varphi		\nonumber\\[4pt]
				&\qquad\quad+\tfrac12\partial_\mu\varphi\partial^\mu\varphi-12g^2e^{-\frac{\varphi}{\sqrt3}}\Big](d\alpha^2+\sin^2\!\alpha d\beta^2+\mu_1^2D\phi_1^2+\mu_2^2D\phi_2^2+\mu_3^2D\phi_3^2)		\nonumber\\[4pt]
				&\quad+\tfrac12e^{\sqrt3\varphi}\Big[g^{\mu\nu}G_{\mu\nu}+\tfrac1{12}e^{\frac{\varphi}{\sqrt3}}H_{\mu\nu}H^{\mu\nu}+\tfrac34e^{\sqrt3\varphi}\tilde H_{4\mu\nu}\tilde H_4^{\mu\nu}-\sqrt3\,\square\varphi		\nonumber\\[4pt]
				&\qquad\quad+\tfrac12\partial_\mu\varphi\partial^\mu\varphi-20g^2e^{-\frac{\varphi}{\sqrt3}}\Big]Dz^2		\nonumber\\[4pt]
				&\quad+\tfrac1{12g^2}e^{\frac{2\varphi}{\sqrt3}}H_{\mu\nu}H^{\mu\nu}\big(\mu_1^2D\phi_1+\mu_2^2D\phi_2+\mu_3^2D\phi_3\big)^2		\nonumber\\[4pt]
				&\qquad+\tfrac1{2\sqrt3}e^{\frac{4\varphi}{\sqrt3}}H_{\mu\nu}\tilde{H}_{4}^{\mu\nu}\big(\mu_1^2D\phi_1+\mu_2^2D\phi_2+\mu_3^2D\phi_3\big)Dz\,,
\end{align}
\end{subequations}
and the stress-energy tensor has components
\begin{subequations}
\begin{align}
	\hat{T}_{\mu\nu}&=-\tfrac13e^{-\frac{\varphi}{\sqrt3}}\Big[12g^2g_{\mu\nu}-e^{\frac{2\varphi}{\sqrt3}}\Big(H_{\mu\rho}H_{\nu}{}^{\rho}-\tfrac14g_{\mu\nu}H_{\rho\sigma}H^{\rho\sigma}\Big)\Big]\,,	\\[8pt]
	\hat{T}_{\mu m}v^m&=0\,,	\\[8pt]
	\hat{T}_{mn}v^mv^n&=4g^2(d\alpha^2+\sin^2\!\alpha d\beta^2+\mu_1^2D\phi_1^2+\mu_2^2D\phi_2^2+\mu_3^2D\phi_3^2)		\nonumber\\[4pt]
				&\quad-\tfrac1{12}e^{\frac{2\varphi}{\sqrt3}}\Big(48g^2+e^{\frac{2\varphi}{\sqrt3}}H_{\mu\nu}H^{\mu\nu}\Big)Dz^2		\nonumber\\[4pt]
				&\quad+\tfrac1{12g^2}e^{\frac{2\varphi}{\sqrt3}}H_{\mu\nu}H^{\mu\nu}\big(\mu_1^2D\phi_1+\mu_2^2D\phi_2+\mu_3^2D\phi_3\big)^2		\nonumber\\[4pt]
				&\qquad+\tfrac1{6g}e^{\sqrt3\varphi}H_{\mu\nu}(*H)^{\mu\nu}\big(\mu_1^2D\phi_1+\mu_2^2D\phi_2+\mu_3^2D\phi_3\big)Dz\,.
\end{align}
\end{subequations}
It is immediate to verify that the external components of \eqref{eq: EinsteinIIB} are satisfied on the four-dimensional Einstein equations, the mixed components amount to the Maxwell equation for $H$ and Bianchi identity for $H^4$, and the internal components of \eqref{eq: EinsteinIIB} are a combination of the trace of the Einstein equations and the equations of motion for the scalars.


\section{Black Hole and Domain Wall Solutions}		\label{sec: BHDW}

In the previous section, we have described how to embed any solution of the four-dimensional gauged STU models presented in section~\ref{sec: STUsugra} into M-theory or type IIB supergravity. We now switch gears to present new black-hole solutions in these theories, first by considering singular limits of previously known solutions and later by directly solving the equations of motion for a suitable Ansatz.

\subsection{AdS-BH limits}
Setting the axions to zero, the potential \eqref{eq: potso8} for the STU truncation of the ${\rm SO(8)}$ gauging reads
\begin{equation}	\label{eq: potso8nochi}
	V=-8g^2(\cosh\varphi_1+\cosh\varphi_2+\cosh\varphi_3)\,.
\end{equation}
This theory admits four-charge AdS-black hole solutions~\cite{Cvetic:1999xp},
\begin{equation}	\label{eq: 4chargeBH}
	\begin{aligned}
		ds_4^2&=-(H_1H_2H_3H_4)^{-1/2}f\,dt^2+(H_1H_2H_3H_4)^{1/2}\big(f^{-1}dr^2+r^2 d\Omega_{2,\kappa}^2\big)\,,	\\[5pt]
		A^i&=H_i^{-1}\frac{\sqrt{\sigma_{\kappa}+\kappa\,\sinh^2(\sqrt{\sigma_{\kappa}}\beta_i)}}{\sinh(\sqrt{\sigma_{\kappa}}\beta_i)}dt\,,		\\[5pt]
		e^{2\varphi_1}&=\frac{H_1H_2}{H_3H_4}\,,	\qquad
		e^{2\varphi_2}=\frac{H_1H_3}{H_2H_4}\,,	\qquad
		e^{2\varphi_3}=\frac{H_1H_4}{H_2H_3}\,,	
	\end{aligned}
\end{equation}
with 
\begin{equation}
	f=\kappa-\frac{\mu}r+4g^2r^2\,H_1H_2H_3H_4		\qquad\quad\text{and}\qquad\quad
	H_i=1+\frac{\mu}r\Big(\frac1{\sqrt{\sigma_\kappa}} \sinh(\sqrt{\sigma_\kappa}\beta_i)\Big)^2\,.
\end{equation}
Here, $d\Omega^2_{2,\kappa}$ denotes the line element for the unit radius metric on the sphere, torus and hyperboloid for $\kappa=1,0,-1$, respectively given by
\begin{equation}
	d\Omega^2_{2,\kappa}=\begin{cases}
		d\theta^2+\sin^2\!\theta\, d\phi^2,		&	\kappa=+1\,,\\
		d\theta^2+d\phi^2,				&	\kappa=0\,,\\
		d\theta^2+\sinh^2\!\theta\, d\phi^2,	&	\kappa=-1\,,
	\end{cases}
\end{equation}
and we also define
\begin{equation}
	\sigma_\kappa=\lim_{\epsilon\to0^+}\text{sign}(\kappa+\epsilon)=
	\begin{cases}
		+1,	&	\kappa=+1,0\,,\\
		-1,	&	\kappa=-1\,.
	\end{cases}
\end{equation}
These black holes have charges\footnote{Our $\beta_i$ parameters are always taken to be real. These configurations agree with those in \cite{Cvetic:1999xp} if one identifies $\beta_i^{\text{there}}=i\, \beta_i^{\text{here}}$ in the hyperbolic case.}
\begin{equation}
	q_i=
	\frac{\pi\mu}{\sqrt{\sigma_\kappa}} \sinh(\sqrt{\sigma_\kappa}\beta_i)\, \sqrt{\sigma_{\kappa}+\kappa\,\sinh^2(\sqrt{\sigma_{\kappa}}\beta_i)}
	=
	\begin{cases}
		\tfrac{1}2\pi\mu\sinh2\beta_i\,,	& \kappa=+1\,,	\\
		\pi\mu\sinh\beta_i\,,			& \kappa=0\,,			\\
		\tfrac{1}2\pi\mu\sin2\beta_i\,,	& \kappa=-1\,.
	\end{cases}
\end{equation}

Domain-wall limits of these solutions can be constructed by considering rescalings of the metric and vectors and shifts of the dilatons so that the equations of motion remain invariant up to a change in the scalar potential resulting from the loss of some of the terms in \eqref{eq: potso8nochi} after taking a singular limit \cite{Cvetic:1999pu}. Taking
\begin{equation}
	\varphi_i\mapsto\varphi_i+\lambda\,,		\qquad\quad\text{with}\qquad 
	\lambda=\log(\tilde{g}/g)\,,
\end{equation}
the potential \eqref{eq: potso8nochi} becomes
\begin{equation}	\label{eq: potsscaled+++}
	V=-4\tilde{g}^2(e^{\varphi_1}+e^{\varphi_2}+e^{\varphi_3})-4g^2(e^{-\varphi_1}+e^{-\varphi_2}+e^{-\varphi_3})\,,
\end{equation}
and the equations of motion are solved by \eqref{eq: 4chargeBH} with now
\begin{equation}
	\begin{aligned}
		&H_1=\Big(\frac{g}{\tilde{g}}\Big)^2\Big[1+\frac{\mu}r\Big(\frac1{\sqrt{\sigma_\kappa}} \sinh(\sqrt{\sigma_\kappa}\beta_1)\Big)^2\Big]\,,	\\[5pt]
		&H_{i\neq1}=1+\frac{\mu}r\Big(\frac1{\sqrt{\sigma_\kappa}} \sinh(\sqrt{\sigma_\kappa}\beta_{i\neq1})\Big)^2\,,
	\end{aligned}
\end{equation}
and
\begin{equation}
	f=\kappa-\frac{\mu}r+4\tilde{g}^2r^2\,H_1H_2H_3H_4\,.
\end{equation}

In this case, it is possible to take the $g\to0$ limit of the potential \eqref{eq: potsscaled+++} while keeping non-trivial solutions provided that $g\sinh(\sqrt{\sigma_\kappa}\beta_i)$ can be kept constant by sending $\beta_i$ to infinity. Note that this cannot be achieved for the hyperbolic horizon ($\kappa=-1$) case above, and therefore only domain walls with spherical and toroidal horizons can be constructed with this method.

We note that the potential \eqref{eq: potsscaled+++} after the $g\to0$ limit corresponds to the STU sector of a maximal supergravity with $\SO(6)\ltimes\mathbb{R}^{12}$ gauging which is related by duality to the one discussed above. In particular, instead of by the embedding tensor $\Theta_\6$ in \eqref{eq: gaugings}, it is described by
\begin{equation}
	\begin{aligned}
		&\theta_{AB}=0\,,		\qquad
		\xi^{AB}=0\,,			\qquad
		\zeta^A{}_{BCD}=0\,,	\\[5pt]
		\tilde{\zeta}_4{}^{123}=&-\tilde{\zeta}_3{}^{124}=\tilde{\zeta}_2{}^{134}=-\tilde{\zeta}_1{}^{234}=\tilde{\zeta}_5{}^{678}=-\tilde{\zeta}_6{}^{578}=\tilde{g}\,,
	\end{aligned}	
\end{equation}
with only non-trivial components in the $\bm{420}$ representation. Using the same scalings as the ones studied in section~\ref{sec: SO8inS7}, this gauged supergravity can be proved to uplift consistently into an $S^5\times T^2$ configuration of $D=11$ supergravity.

\vspace{8pt}

The potential \eqref{eq: potso6so2} can be similarly obtained by taking
\begin{equation}
	\varphi_1\mapsto\varphi_1-\lambda\,,		\qquad
	\varphi_{2,3}\mapsto\varphi_{2,3}+\lambda\,,		\qquad\text{with}\qquad 
	\lambda=\log(\tilde{g}/g)\,,
\end{equation}
so that  \eqref{eq: potso8nochi} becomes
\begin{equation}	\label{eq: potsscaled-++}
	V=-4\tilde{g}^2(e^{-\varphi_1}+e^{\varphi_2}+e^{\varphi_3})-4g^2(e^{\varphi_1}+e^{-\varphi_2}+e^{-\varphi_3})\,.
\end{equation}
However, in this case the rescaled solution is \eqref{eq: 4chargeBH} with
\begin{equation}
	\begin{aligned}
		&H_2=\Big(\frac{g}{\tilde{g}}\Big)^{-1}\Big[1+\frac{\mu}r\Big(\frac1{\sqrt{\sigma_\kappa}} \sinh(\sqrt{\sigma_\kappa}\beta_2)\Big)^2\Big]		\qquad\text{and}\\[5pt]
		&H_{i\neq2}=\Big(\frac{g}{\tilde{g}}\Big)\Big[1+\frac{\mu}r\Big(\frac1{\sqrt{\sigma_\kappa}} \sinh(\sqrt{\sigma_\kappa}\beta_{i\neq2})\Big)^2\Big]\,,
	\end{aligned}
\end{equation}
so that the $g\to0$ limit is ill-defined.

\subsection{DW-Ans\"atze}

The fact that domain-wall solutions for the $\SO(6)\ltimes\mathbb{R}^{12}$ and $[\SO(6)\times\SO(2)]\ltimes\mathbb{R}^{12}$ theories cannot be constructed through singular limits of the AdS-black holes in \cite{Cvetic:1999xp} does not mean that such solutions do not exist.
To find them, we consider the simpler truncation in \eqref{eq: 3+1truncation} and the Ansatz 
\begin{equation}	\label{eq: DWAnsatz}
	\begin{aligned}
		ds_4^2=-e^{2u}dt^2+e^{-2u}dr^2+L^2e^{2A}d\Omega^2_{2,\kappa}\,,	\qquad
		A_\1=v\, dt\,,	\qquad
		A_{\1}^4=w\, dt\,,
	\end{aligned}
\end{equation}
with $\varphi, u, v, w$ and $A$ being functions on $r$ only, and $L$ a constant radius. The equations of motion stemming from \eqref{eq: 111Lagrangian} reduce~to
\begin{equation}	\label{eq: Ansatzeqns}
	\begin{aligned}
		&\varphi''+2(u'+A')\varphi'+\tfrac12 a\, e^{a\varphi-2u}(v')^2-\tfrac1{2a}\, e^{-\varphi/a-2u}(w')^2-12 a\,g^2 e^{-a\varphi-2u}=0\,,	\\[5pt]
		&(e^{2A+a\varphi}v')'=0\,,		\\[5pt]
		&(e^{2A-\varphi/a}w')'=0\,,		\\[5pt]
		&A''+A'+\tfrac14(\varphi')^2=0\,,		\\[5pt]
		&A''+2A'(u'+A')+\tfrac14 e^{a\varphi-2u}(v')^2+\tfrac14 e^{-\varphi/a-2u}(w')^2-12 g^2 e^{-a\varphi-2u}-\tfrac{\kappa}{L^2}\,e^{-2A-2u}=0\,,		\\[5pt]
		&u''+2u'(u'+A')-\tfrac14 e^{a\varphi-2u}(v')^2-\tfrac14 e^{-\varphi/a-2u}(w')^2-12 g^2 e^{-a\varphi-2u}=0\,,	
	\end{aligned}
\end{equation}
with primes denoting derivatives with respect to $r$. For any $a$ in the range $-\sqrt3<a<\sqrt3$, this system of equations admits a solution in the $\kappa=0$ case, which reads
\begin{equation}	\label{eq: DWsolna}
	\begin{aligned}
		&ds_4^2=-f\,dt^2+f^{-1}dr^2+L^2\Big(\frac{r}{r_0}\Big)^{2/(1+a^2)}d\Omega^2_{2,0}\,,	\\[5pt]
		&A_\1=\frac\lambda{r}\, dt\,,	\qquad\quad
		A_\1^4=0\,,	\\[5pt]
		&e^\varphi=\Big(\frac{r}{r_0}\Big)^{2a/(1+a^2)}\,,
	\end{aligned}
\end{equation}
with
\begin{equation}	
	f=\frac{1+a^2}4\left[\frac{24g^2r_0^2(1+a^2)}{3-a^2}\Big(\frac{r}{r_0}\Big)^{2/(1+a^2)}+\frac{\lambda^2}{r_0^2}\Big(\frac{r_0}{r}\Big)^{2/(1+a^2)}\right]-\frac{\mu}{r_0} \Big(\frac{r_0}{r}\Big)^{(1-a^2)/(1+a^2)}\,,
\end{equation}
for $r_0$ a length parameter and $\lambda$ and $\mu$ constants encoding the charge and mass of the solution. This class of solutions recovers (A.6) of \cite{Cvetic:1999pu} for $a=1$, while for $a=1/\sqrt3$ it reduces to
\begin{align}	\label{eq: DWsolna}
		&ds_4^2=-\sqrt{\frac{r}{r_0}}\left[4g^2 r_0 r+\frac{\lambda^2}{3r^2}-\frac{\mu}{r}\right]\,dt^2+\sqrt{\frac{r_0}{r}}\left[4g^2 r_0 r+\frac{\lambda^2}{3r^2}-\frac{\mu}{r}\right]^{-1}dr^2+L^2\Big(\frac{r}{r_0}\Big)^{3/2}d\Omega^2_{2,0}\,,	\nonumber\\[5pt]
		&A_\1=\frac\lambda{r}\, dt\,,		\qquad\quad
		A_\1^4=0\,,	\nonumber\\[5pt]
		&e^\varphi=\Big(\frac{r}{r_0}\Big)^{\sqrt3/2}\,,	
\end{align}
which describes a charged black hole with domain-wall asymptotics. At $r=0$, this solution has a curvature singularity for all possible values of the charge and mass. When
\begin{equation}	\label{eq: extremality}
	\mu^3>3 g^2 \lambda^4 r_0\,,
\end{equation}
the singularity is covered by two horizons sitting at
\begin{equation}
	r_\pm=\sqrt{\frac{\mu}{3 g^2 r_0}} \cos \left[\frac{1}{3} \arccos\left(-g \lambda ^2 \sqrt{\frac{3 r_0}{\mu ^3}}\right)-\frac{\pi }{3}\pm\frac{\pi }{3}\right]
\end{equation}
that coalesce when the bound \eqref{eq: extremality} is saturated. When $\mu^3<3 g^2 \lambda^4 r_0$, the solution displays a naked singularity.

For this choice of dilaton coupling, we can embed the solution in ten dimensions using \eqref{eq: metricIIB3+1}--\eqref{eq: fiveformIIB3+1}
with the tensor hierarchy fields in \eqref{eq: TH3+1}, which become
\begin{equation}
	\begin{aligned}
		\tilde{H}_{\2\,1,2,3}&=-\frac{\lambda L^2}{\sqrt3r_0^2}d\theta\wedge d\phi\,,		\qquad\quad 
		\tilde{H}_{\2\,4}=0\,,		\\[5pt]
		H_{\3\,1,2,3}&=-\frac13H_{\3\,4}=\frac{L^2}{4r}\Big(\frac{r}{r_0}\Big)^2\left[4g^2 r_0 r+\frac{\lambda^2}{3r^2}-\frac{\mu}{r}\right]dt\wedge d\theta\wedge d\phi\,,			\\[5pt]
		H_{\4\,1,2,3}&=4gL^2\, \Big(\frac{r}{r_0}\Big)\,dt\wedge dr\wedge d\theta\wedge d\phi\,,			\qquad\, 
		H_{\4\,4}=6gL^2\, \Big(\frac{r}{r_0}\Big)^{2}\,dt\wedge dr\wedge d\theta\wedge d\phi\,,				\\[5pt]
		\tilde{H}_{\4\,1,2,3}&=2mL^2\, \Big(\frac{r}{r_0}\Big)\,dt\wedge dr\wedge d\theta\wedge d\phi\,,	\quad\enspace\ 
		\tilde{H}_{\4\,4}=0\,.			
	\end{aligned}
\end{equation}

Other solutions to the equations \eqref{eq: Ansatzeqns} can be found for specific values of the dilaton coupling $a$. If we commit to the choice $a=1/\sqrt3$ corresponding to the STU model, a class of solutions with spherical horizon is given by
\begin{equation}	\label{eq: DWsolnb}
	\begin{aligned}
		&ds_4^2=-f\,dt^2+f^{-1}dr^2+\lambda^2\Big(\frac{r}{r_0}\Big)^{3/2}d\Omega^2_{2,1}\,,	\qquad\text{with}\quad
		f=\sqrt{\frac{r}{r_0}}\left[4g^2r_0r+\frac{r_0^2}{\lambda^2}-\frac\mu{r}\right]\,,	\\[5pt]
		&A_\1=0\,,		\qquad\qquad
		A^4_\1=\frac{r}\lambda\, dt\,,		\qquad\qquad
		e^\varphi=\Big(\frac{r}{r_0}\Big)^{\sqrt3/2}\,.
	\end{aligned}
\end{equation}
Note that this solution only exists for non-vanishing electric charge, and for $\mu>0$ it has a single horizon at radius
\begin{equation}
	r_H=\frac{r_0}{8g^2\lambda^2}\left[\sqrt{1+\frac{16g^2\lambda^4\mu}{r_0^3}}-1\right]\,.
\end{equation}
For this solution, the type IIB uplift is given by \eqref{eq: metricIIB3+1}--\eqref{eq: fiveformIIB3+1}
with tensor hierarchy fields
\begin{equation}
	\begin{aligned}
		\tilde{H}_{\2\,1,2,3}&=0\,,		\qquad\quad 
		\tilde{H}_{\2\,4}=\lambda\sin\theta\, d\theta\wedge d\phi\,,		\\[5pt]
		H_{\3\,1,2,3}&=-\frac13H_{\3\,4}=\frac{\lambda^2}{4r}\Big(\frac{r}{r_0}\Big)^2\left[4g^2 r_0 r+\frac{r_0^2}{\lambda^2}-\frac{\mu}{r}\right]\sin\theta\,dt\wedge d\theta\wedge d\phi\,,			\\[5pt]
		H_{\4\,1,2,3}&=4g\lambda^2\, \Big(\frac{r}{r_0}\Big)\,\sin\theta\,dt\wedge dr\wedge d\theta\wedge d\phi\,,			\qquad\, 
		H_{\4\,4}=6g\lambda^2\, \Big(\frac{r}{r_0}\Big)^{2}\,\sin\theta\,dt\wedge dr\wedge d\theta\wedge d\phi\,,				\\[5pt]
		\tilde{H}_{\4\,1,2,3}&=2m\lambda^2\, \Big(\frac{r}{r_0}\Big)\,\sin\theta\,dt\wedge dr\wedge d\theta\wedge d\phi\,,	\quad\enspace\ 
		\tilde{H}_{\4\,4}=0\,.			
	\end{aligned}
\end{equation}
%


\section{Discussion}

In this paper, we have shown that singular limits of gauged supergravities can offer insights not only into how to construct new solutions in the resulting gauged supergravity, but also on its consistent uplift to higher dimensions if the resulting gauged supergravity can itself be obtained as a consistent truncation. We have exemplified this idea relating the STU sector of the electrically gauged SO(8) supergravity to the STU sector of the CSO theory with the $\SO(6)\ltimes\mathbb{R}^{12}$ gauging, and we have used the known consistent uplift of the former theory into M-theory on $S^7$ to construct an embedding of the latter into type IIB supergravity on $S^5\times S^1$. It would be interesting to investigate if singular limits such as the ones we studied can be used to relate gauged supergravities with larger field contents.

To describe the uplifts of these gauged supergravities, we have employed techniques that exploit the formal duality-covariance of the embedding tensor formulation of the gauging. In particular, 
apart from making use of the generalised Scherk-Schwarz factorisation of the higher-dimensional fields expressed in the language of Exceptional Field Theory, we
have recast all the contributions of $p$-form fields in $D=4$ in terms of the four-dimensional tensor hierarchy (suitably restricted to the STU sector). This has provided a simple way of circumventing some complicated dualisations in ExFT involving the internal metric. 
To illustrate the power of this technique, we have explicitly checked that the higher-dimensional equations of motion follow from the four-dimensional ones, and constructed new families of charged black hole solutions in $4d$. These black holes involve non-trivial scalar profiles, and their asymptotics do not approach an AdS solution, but a domain wall. 

A close cousin of the dyonic CSO theory we have considered comes equipped with an $[\SO(6)\times\SO(1,1)]\ltimes\mathbb{R}^{12}$ gauge group. This theory has been shown to uplift into type IIB supergravity on $S^5\times S^1$ with a non-geometric patching of the circle \cite{Inverso:2016eet}. This theory possesses a very rich structure of AdS vacua \cite{Guarino:2020gfe}, both supersymmetric and non-supersymmetric, including continuous families realising a holographic conformal manifold \cite{Bobev:2021yya,Bobev:2021rtg,Giambrone:2021wsm,Giambrone:2021zvp,Cesaro:2021tna,Cesaro:2022mbu}. It will be interesting to extend our results to describe more explicitly the uplift of this gauging, and to construct in this way ten-dimensional solutions arising from non-trivial profiles of the fields in the consistent truncation. These new solutions may play an important r\^ole in understanding the holography of the ${\rm T}[{\rm U}(N)]$ theories and the $J$-folds of $\mathcal{N}=4$ SYM that are conjectured to be dual to this solution \cite{Assel:2018vtq}.


\section*{Acknowledgements}

We would like to thank Camille Eloy, Henning Samtleben and Oscar Varela for interesting discussions and comments. HZ would like to thank the Harvard Physics Department and CCPP at New York University for kind hospitality during the later stages of this project. GL is supported by endowment funds from the Mitchell Family Foundation. CNP is supported in part by DOE grant DE-SC0010813.

\appendix

\section{STU Truncation} \label{app: Cartans}

In its $\SL(8,\mathbb{R})$ basis, the generators of E$_{7(7)}$ can be split as $t_\alpha=\{t_A{}^B,\ t_{[ABCD]}\}$, with the index $A=1,\dots,8$ labelling the fundamental of $\SL(8,\mathbb{R})$ and $t_A{}^A=0$. The gaugings considered are described by an embedding tensor $\Theta_M{}^\alpha$ with non-trivial components
\begin{equation}	\label{eq: embtensor912}
	\Theta_{AB}{}^{C}{}_{D}=2\delta_{[A}{}^C\theta_{B]D}\,,	\qquad\quad
	\Theta^{ABC}{}_{D}=2\delta_D{}^{[A} \xi^{B]C}\,,
\end{equation}
where $\theta_{AB}$ and $\xi^{AB}$ respectively belong to the $\bm{36'}$ and $\bm{36}$ of $\SL(8,\mathbb{R})$. This embedding tensor determines the covariant derivatives as
\begin{equation}
	\begin{aligned}
		D&=d+A^{M}\Theta_M{}^\alpha t_\alpha	\\[5pt]
		&=d+A^{AB}\,\theta_{BC}\,t_A{}^C+\tilde{A}_{AB}\,\xi^{BC}\,t_C{}^A\,.
	\end{aligned}
\end{equation}
and appears in the potential and mass matrices through the generators $X_{MN}{}^P=\Theta_M{}^\alpha(t_\alpha)_N{}^P$.

For these gaugings, the gauge group $G$ is embedded in ${\rm E}_{7(7)}$ as
\begin{equation}	\label{eq: groupinclusions}
	G\subset\SL(8,\mathbb{R})\subset {\rm E}_{7(7)}\,.
\end{equation}
The $\mathcal{H}={\rm SO}(2)^4$ Cartan subalgebra of the gaugings discussed in the main text can be taken to be\footnote{For the $\SO(6)\ltimes\mathbb{R}^{12}$ gauging, $T_{78}$ strictly speaking corresponds to a global symmetry commuting with the gauge group, following the $m\rightarrow0$ limit.}
\begin{equation}	\label{eq: T1234}
	T_{12}=t_1{}^2-t_2{}^1\,,	\qquad
	T_{34}=t_3{}^4-t_4{}^3\,,	\qquad
	T_{56}=t_5{}^6-t_6{}^5\,,	\qquad
	T_{78}=t_7{}^8-t_8{}^7\,,
\end{equation}
and its commutant within ${\rm E}_{7(7)}$ is
\begin{equation}	\label{eq: commutant}
	{\rm Comm}_{\mathcal{H}}\,{\rm E}_{7(7)}=\SL(2,\mathbb{R})_1\times\SL(2,\mathbb{R})_2\times\SL(2,\mathbb{R})_3\,.
\end{equation}
Each factor is generated by $\{h_i, e_i, f_i\}$, which are given in terms of ${\rm E}_{7(7)}$ generators as
\begin{equation}	\label{eq: sl2}
	\begin{aligned}
		h_1&=\tfrac14\big(t_1{}^1+t_2{}^2+t_3{}^3+t_4{}^4-t_5{}^5-t_6{}^6-t_7{}^7-t_8{}^8\big)\,,	\\
		&\qquad
		e_1=12\,t_{1234}\,,	\qquad\qquad
		f_1=12\,t_{5678}\,,	\\[7pt]
		h_2&=\tfrac14\big(t_1{}^1+t_2{}^2-t_3{}^3-t_4{}^4-t_5{}^5-t_6{}^6+t_7{}^7+t_8{}^8\big)\,,	\\
		&\qquad
		e_2=12\,t_{1278}\,,	\qquad\qquad
		f_2=12\,t_{3456}\,,	\\[7pt]
		h_3&=\tfrac14\big(-t_1{}^1-t_2{}^2+t_3{}^3+t_4{}^4-t_5{}^5-t_6{}^6+t_7{}^7+t_8{}^8\big)\,,	\\
		&\qquad
		e_3=12\,t_{3478}\,,	\qquad\qquad
		f_3=12\,t_{1256}\,,	
	\end{aligned}
\end{equation}
with non-vanishing brackets
\begin{equation}	\label{eq: sl2alg}
	[h_i,\, e_j]=e_i\,\delta_{ij}\,,	\qquad
	[h_i,\, f_j]=-f_i\,\delta_{ij}\,,	\qquad
	[e_i,\, f_j]=2h_i\,\delta_{ij}\,.
\end{equation}
The representative of the scalar manifold \eqref{eq: scalarmanif} is taken to be
\begin{equation}	\label{eq: cosetV}
	\mathcal{V}=e^{\chi_1 e_1+\chi_2 e_2+\chi_3 e_3}\, e^{-\varphi_1 h_1-\varphi_2 h_2-\varphi_3 h_3}\,,
\end{equation}
which leads to \eqref{eq: STULagrangian} through the symmetric matrix $M_{MN}=(\mathcal{V}\mathcal{V}^T)_{MN}$ and its inverse as 
\begin{equation}	\label{eq: NLSMwithM}
	\mathcal{L}_{\rm NLSM}=-\tfrac1{48}dM_{MN}\wedge*dM^{MN}\,.
\end{equation}
The symmetry group \eqref{eq: commutant} is then realised through the Killing vectors
\begin{equation}	\label{eq: NLSMKilling}
	k[h_i]=2\partial_{\varphi_i}-2\chi_i\partial_{\chi_i}\,,	\qquad\quad
	k[e_i]=\partial_{\chi_i}\,,						\qquad\quad
	k[f_i]=2\chi_i\partial_{\varphi_i}+(e^{-2\varphi_i}-\chi_i^2)\partial_{\chi_i}\,,
\end{equation}
which close into \eqref{eq: sl2alg} under the Lie bracket.

There are four electric vectors $A^{\rm a}$ and four magnetic counterparts $\tilde{A}_{\rm a}$ in the $\mathcal{H}$-invariant truncation of the maximal theory. In terms of the original $\bm{28}\oplus\bm{28}'$ of the $\mathcal{N}=8$ theory, they are given by
\begin{equation}	\label{eq: TensorHierarchyA}
	\begin{tabular}{c@{\hskip 27pt}c@{\hskip 27pt}c@{\hskip 27pt}c}
		$A^1=A^{12}\,,$			&	$A^2=A^{34}\,,$			&	$A^3=A^{56}\,,$			&	$A^4=A^{78}\,,$				\\[5pt]
		$\tilde{A}_1=\tilde{A}_{12}\,,$	&	$\tilde{A}_2=\tilde{A}_{34}\,,$	&	$\tilde{A}_3=\tilde{A}_{56}\,,$	&	$\tilde{A}_4=\tilde{A}_{78}\,.$	
	\end{tabular}
\end{equation}
These identifications allow one to extract the non-minimal gauge couplings from $M$ through the block decomposition
\begin{equation}	\label{eq: RIinM} 
	M=\begin{pmatrix}
		-\mathcal{I}-\mathcal{R}\mathcal{I}^{-1}\mathcal{R}	&	\mathcal{R}\mathcal{I}^{-1}	\\
		\mathcal{I}^{-1}\mathcal{R}					&	-\mathcal{I}^{-1}
	\end{pmatrix}\,,
\end{equation}
and the index restriction $A^M=(A^{[AB]},\tilde{A}_{[AB]})\to(A^{\rm a},\tilde{A}_{\rm a})$ in \eqref{eq: TensorHierarchyA}.

Similarly, the preserved two-forms in \eqref{eq: THcontentFull} are related to the ${\rm SL}(8,\mathbb{R})$ objects through
\begin{gather}	
\label{eq: TensorHierarchyB}
	B_1=B_1{}^{1}=B_2{}^{2}\,,	\hspace{27pt}	B_2=B_3{}^{3}=B_4{}^{4}\,,	\hspace{27pt}	B_3=B_5{}^{5}=B_6{}^{6}\,,	\hspace{27pt}	B_4=B_7{}^{7}=B_8{}^{8}\,,	\nonumber\\[5pt]
	B'_1=B_1{}^{2}=-B_2{}^{1}\,,	\hspace{18pt}	B'_2=B_3{}^{4}=-B_4{}^{3}\,,	\hspace{18pt}	B'_3=B_5{}^{6}=-B_6{}^{5}\,,	\hspace{18pt}	B'_4=B_7{}^{8}=-B_8{}^{7}\,,	\nonumber\\[5pt]
	B_{12}=B_{1234}\,,			\hspace{27pt}	B_{13}=B_{1256}\,,			\hspace{27pt}	B_{14}=B_{1278}\,,				\nonumber	\\[5pt]
	B_{23}=B_{3456}\,,			\hspace{27pt}	B_{24}=B_{3478}\,,			\hspace{27pt}	B_{34}=B_{5678}\,,			
\end{gather}
and the three-forms as
\begin{equation}	\label{eq: TensorHierarchyC36}
	\begin{tabular}{c@{\hskip 27pt}c@{\hskip 27pt}c@{\hskip 27pt}c}
		$C_1=C^{11}=C^{22}\,,$			&	$C_2=C^{33}=C^{44}\,,$			&	$C_3=C^{55}=C^{66}\,,$			&	$C_4=C^{77}=C^{88}\,,$	\\[5pt]
		$\tilde{C}_1=\tilde C_{11}=\tilde C_{22}\,,$		&	$\tilde{C}_2=\tilde C_{33}=\tilde C_{44}\,,$		&	$\tilde{C}_3=\tilde C_{55}=\tilde C_{66}\,,$		&	$\tilde{C}_4=\tilde C_{77}=\tilde C_{88}\,,$
	\end{tabular}
\end{equation}
for those in the $\bm{36}\oplus\bm{36'}$ of ${\rm SL}(8,\mathbb{R})$, and those in the $\bm{420}\oplus\bm{420'}$ as
\begin{equation}	\label{eq: TensorHierarchyC420}
	\begin{gathered}
	C_{12}=C_{1}{}^{234}=-C_{2}{}^{134}\,,	\hspace{27pt}	C_{13}=C_{1}{}^{256}=-C_{2}{}^{145}\,,	\hspace{27pt}	C_{14}=C_{1}{}^{278}=-C_{2}{}^{178}\,,	\\[5pt]
	C_{21}=C_{3}{}^{412}=-C_{4}{}^{312}\,,	\hspace{27pt}	C_{23}=C_{3}{}^{456}=-C_{4}{}^{356}\,,	\hspace{27pt}	C_{24}=C_{3}{}^{478}=-C_{4}{}^{378}\,,	\\[5pt]
	\tilde{C}_{12}=\tilde{C}^{1}{}_{234}=-\tilde{C}^{2}{}_{134}\,,	\hspace{27pt}	\tilde{C}_{13}=\tilde{C}^{1}{}_{256}=-\tilde{C}^{2}{}_{145}\,,	\hspace{27pt}	\tilde{C}_{14}=\tilde{C}^{1}{}_{278}=-\tilde{C}^{2}{}_{178}\,,	\\[5pt]
	\tilde{C}_{21}=\tilde{C}^{3}{}_{412}=-\tilde{C}^{4}{}_{312}\,,	\hspace{27pt}	\tilde{C}_{23}=\tilde{C}^{3}{}_{456}=-\tilde{C}^{4}{}_{356}\,,	\hspace{27pt}	\tilde{C}_{24}=\tilde{C}^{3}{}_{478}=-\tilde{C}^{4}{}_{378}\,,	\\[5pt]
	etc\,.		\\[-12pt]
	\end{gathered}
\end{equation}\\[-8pt]
%


\section{A $D=9$ Detour} \label{app: conventions}

\subsection{Reduction from $D=11$ to $D=9$}

The relation between the eleven-dimensional and type IIA fields in Einstein frame reads
\begin{equation}	\label{eq: 11Ansatz}
	\begin{aligned}
		d\hat{s}_{11}^2 &= e^{-\frac{1}{6}\phi_A} ds_{\rm IIA}^2 + e^{\frac{4}{3}\phi_A}(dz_2+\mathcal{A}')^2\,,	\\[5pt]
		\hat{F}_{\4} &= F_{\4  \rm IIA} + H_{\3 \rm IIA}\wedge (dz_2+\mathcal{A}')\,. 
	\end{aligned}
\end{equation}
Further reducing type IIA on another circle, we get maximal supergravity in $D=9$. The reduction Ansatz is
\begin{equation}
	\begin{aligned}
		ds_{\rm IIA}^2&= e^{-\frac{1}{2\sqrt{7}}\varphi_A}ds_9^2 + e^{\frac{\sqrt{7}}{2}\varphi_A} \big(dz_1+ \mathcal{A}^{\rm KK}\big)^2\,,	\\[5pt]
		H_{\3 \rm IIA} &=  H^A_\3 +  H^A_\2 \wedge \big(dz_1+ \mathcal{A}^{\rm KK}\big)	\,,	\\[5pt]
		\mathcal{F}'_{\2 \rm IIA}&=\mathcal{F}^A_{\2}+\mathcal{F}^A_{\1}\wedge \big(dz_1+ \mathcal{A}^{\rm KK}\big)	\,,	\\[5pt]
		F_{\4 \rm IIA} &=  F^A_\4 +  F^A_\3\wedge \big(dz_1+ \mathcal{A}^{\rm KK}\big)	\,,	
	\end{aligned}
\end{equation}
with $\mathcal{F}'_{\2 \rm IIA}= d\mathcal{A}'$ and the $D=9$ field strengths given by
\begin{align}	\label{eq: IIApots}
		&\mathcal{F}_\1^A=dA_\0^{\rm RR}	\,,										\nonumber\\[5pt]
		&\mathcal{F}_\2^A=dA_\1^{\rm RR}-dA_\0^{\rm RR}\wedge\mathcal{A}_\1^{\rm KK}\,,	\qquad\quad
		H_\2^A=dA_\1^{\rm NS}\,,													\nonumber\\[5pt]
		&H_\3^A=dA_\2^{\rm NS}-dA_\1^{\rm NS}\wedge \mathcal{A}_\1^{\rm KK}\,,			\qquad\quad\ 
		F_\3^A=dA_\2^{\rm RR}+A_\1^{\rm NS}\wedge dA_\1^{\rm RR}-A_\0^{\rm RR}\, dA_\2^{\rm NS}\,,\nonumber\\[5pt]
		&F_\4^A=dA_\3^{\rm RR}+A_\2^{\rm NS}\wedge dA_\1^{\rm RR}-(dA_\2^{\rm RR}-A_\0^{\rm RR}\,dA_\2^{\rm NS}-A_\1^{\rm NS}\wedge dA_\1^{\rm RR})\wedge \mathcal{A}_\1^{\rm KK}\,.	
\end{align}
Comparing with \eqref{eq: scaledMetric} and \eqref{eq: scaled4formTotal}, we find
\begin{gather}	\label{eq: IIAsoln}
	e^{\phi_{\rm IIA}} = \frac{\Xi_2^{1/4}}{\sqrt{H}}\,,\quad \enspace 
	e^{-\tfrac{2\sqrt7}3\varphi_A}= \frac{H}{\sqrt{\Xi_2}}\,,		\quad\enspace
	A_\1^{\rm NS}= \mathcal{A}\,,									\nonumber\\[5pt]
	A_\0^{\rm RR}=\mathcal{A}^{\rm KK}=A^{\rm RR}_{\1}=A^{\rm NS}_{\2}=A_\2^{\rm RR} = 0\,,	\quad\enspace
	dA_\3^{\rm RR}=X_\4\,,
\end{gather}
with $\mathcal{A}$ in \eqref{eq: IIBKKvector}, and 
\begin{equation}
	\begin{aligned}
		ds_{9}^2& =H^{-6/7}\Xi_2^{-4/7}\Big[H\Xi_2\, ds_4^2-(\mu_1^2 b_2 D\phi_1 +\mu_2^2b_3D\phi_2-\mu_3^2 b_1 D\phi_3)^2		\\[5pt]
		&\qquad+H\big(Y_2^2(d\mu_1^2+\mu_1^2D\phi_1^2)+Y_3^2(d\mu_2^2+\mu_2^2D\phi_2^2)+\tilde{Y}_1^2(d\mu_3^2+\mu_3^2D\phi_3^2)\big)\Big]\,.
	\end{aligned}
\end{equation}

\subsection{Type IIB reduction to $9$ dimensions}

The same nine-dimensional theory can be obtained starting from type IIB supergravity. In this case, the Kaluza-Klein Ansatz reads

\begin{equation}
	\begin{aligned}
		ds_{\rm IIB}^2&= e^{-\frac{1}{2\sqrt{7}}\varphi_B}ds_9^2 + e^{\frac{\sqrt{7}}{2}\varphi_B} (dz_1+ \mathcal{B}^{\rm KK})^2\,,		\\[5pt]
		F_{\5 \rm IIB} &=  (1+*_{10})[F^{\rm B}_\4\wedge dz_1]	\,,		\\[5pt]
		F_{\3 \rm IIB} &=  F^{\rm B}_\3 +  F^{\rm B}_\2 \wedge (dz_1+ \mathcal{B}^{\rm KK})	\,,	\\[5pt]
		H_{\3 \rm IIB} &=  H^{\rm B}_\3 +  H^{\rm B}_\2 \wedge (dz_1+ \mathcal{B}^{\rm KK})	\,,
	\end{aligned}
\end{equation}
and the field strengths can be given as \cite{Lavrinenko:1999xi}
\begin{equation}	\label{eq: IIBpots}
	\begin{aligned}
		&H^{\rm B}_\2=dB^{\rm NS}_\1	\,,	\qquad
		F^{\rm B}_\2=dB^{\rm RR}_\1-B^{\rm NS}_\1\wedge d\chi	\,,	\\[5pt]
		&H^{\rm B}_\3=dB^{\rm NS}_\2-dB^{\rm NS}_\1\wedge \mathcal{B}^{\rm KK}_\1	\,,	\\[5pt]
		&F^{\rm B}_\3=dB^{\rm RR}_\2+B^{\rm NS}_\2\wedge d\chi-(dB^{\rm RR}_\1-B^{\rm NS}_\1\wedge d\chi)\wedge \mathcal{B}^{\rm KK}_\1	\,,	\\[5pt]
		&F^{\rm B}_\4=dB^{\rm RR}_\3+B^{\rm RR}_\1\wedge dB^{\rm NS}_\2-B^{\rm NS}_\1\wedge dB^{\rm RR}_\2-B^{\rm NS}_\1\wedge B^{\rm NS}_\2\wedge d\chi\,.
	\end{aligned}
\end{equation}

In this context, T-duality between the two type II theories stems from the uniqueness of maximal $D=9$ supergravity. The relation between IIA and IIB fields can thus be obtained by relating the nine-dimensional scalars, metric and potentials in \eqref{eq: IIApots} and \eqref{eq: IIBpots}.
Using the rules in \cite{Lavrinenko:1999xi} (disregarding the doubled fields there) on \eqref{eq: IIAsoln}, we find that the 9$d$ metrics are equal, the dilatons are given by
\begin{equation}	\label{eq: TdualScalars}
	e^{\phi_B}=1\,,	\qquad
	e^{\varphi_B}=H^{2/\sqrt7}\,\Xi_2^{-1/\sqrt7}\,,
\end{equation}
and the only non-vanishing $p$-forms are
\begin{align}	\label{eq: TdualForms}
	B_\1^{\rm KK}=\mathcal{A}\,,	\qquad
	B_\3^{\rm RR}=A_\3^{\rm RR}\,.
\end{align}
The total type IIB configuration is given in \eqref{eq: IIBfrom11D} in the main text.

\bibliography{references}

\end{document}